%% file: main.tex
\documentclass{jfm}
\usepackage{graphicx}
\usepackage{epstopdf, epsfig}
\usepackage{color}
\usepackage{amsmath,amssymb,amsfonts,bm}
\usepackage{mathtools}

\usepackage[english]{babel}
\usepackage[utf8]{inputenc} 
\usepackage[T1]{fontenc}
 
\usepackage{hyperref}
\hypersetup{
     colorlinks=true,
    linkcolor=blue,
    filecolor=magenta,      
    urlcolor=blue,
    citecolor=blue,
}

\input{commands}

\shorttitle{Mobility of a spheroid near an elastic interface}
\shortauthor{A. Daddi-Moussa-Ider, M. Lisicki and S. Gekle}
\title{Mobility of an axisymmetric particle near an elastic interface}

\author{Abdallah Daddi-Moussa-Ider\aff{1}\corresp{\email{abdallah.daddi-moussa-ider@uni-bayreuth.de}}\footnotemark[2], Maciej Lisicki\aff{2,3}\footnotemark[2]\footnotetext[2]{These authors contributed equally to this work.}, Stephan Gekle\aff{1}} 

\affiliation{\aff{1}Biofluid Simulation and Modeling, Fachbereich Physik, Universit\"at Bayreuth, Universit\"{a}tsstra{\ss}e 30, Bayreuth 95440, Germany.
\aff{2}Department of Applied Mathematics and Theoretical Physics, Wilberforce Rd, Cambridge CB3 0WA, United Kingdom.
\aff{3}Institute of Theoretical Physics, Faculty of Physics, University of Warsaw, Pasteura 5, 02-093 Warsaw, Poland.}

\begin{document}

\maketitle

\begin{abstract}
Using a fully analytical theory, we compute the leading order corrections to the translational, rotational and translation-rotation coupling mobilities of an arbitrary axisymmetric particle immersed in a Newtonian fluid moving near an elastic cell membrane that exhibits resistance towards stretching and bending.
The frequency-dependent mobility corrections are expressed as general relations involving separately the particle's shape-dependent bulk mobility and the shape-independent parameters such as the membrane-particle distance, the particle orientation and the characteristic frequencies associated with shearing and bending of the membrane.
This makes the equations applicable to an arbitrary-shaped axisymmetric particle provided that its bulk mobilities are known, either analytically or numerically.
For a spheroidal particle, these general relations reduce to simple expressions in terms of the particle's eccentricity.
We find that the corrections to the translation-rotation coupling mobility are primarily determined by bending, whereas shearing manifests itself in a more pronounced way in the rotational mobility.
We demonstrate the validity of the analytical approximations by a detailed comparison with boundary integral simulations of a truly extended spheroidal particle.
They are found to be in a good agreement over the whole range of applied frequencies.
\end{abstract}


Published as: J. Fluid Mech. \textbf{811}, 210-233 (2017), \href{https://doi.org/10.1017/jfm.2016.739}{doi:10.1017/jfm.2016.739}

\section{Introduction}
Hydrodynamic interactions between nanoparticles and cell membranes play an important role in many medical and biological applications.
Prime examples are drug delivery and targeting via nanocarriers which release the active agent in disease sites such as tumours or inflammation areas \citep{naahidi13, al-obaidi15, liu16}.
During navigation through the blood stream, but especially during uptake by a living cell via endocytosis \citep{Doherty_2009, meinel14, AgudoCanalejo_2015}, nanoparticles frequently come into close contact with cell membranes which alter their hydrodynamic mobilities in a complex fashion.

Over the last few decades, considerable research effort has been devoted to study the motion of particles in the vicinity of interfaces. 
The particularly simple example of a solid spherical particle has been extensively studied theoretically 
near a rigid no-slip wall \citep{lorentz07, brenner61, goldman67a, goldman67b, cichocki98, swan07, franosch09, happel12}, 
an interface separating two immiscible liquids \citep{lee79, bickel06, bickel07, wanggm09, blawz10, bickel14}, 
an interface with partial-slip \citep{lauga05, felderhof12} 
and a membrane with surface elasticity \citep{felderhof06b, felderhof06, shlomovitz13, shlomovitz14, salez15, daddi16, daddi16b, daddi16c, saintyves16}. 
Elastic membranes stand apart from both liquid-solid and liquid-liquid interfaces, since the elasticity of the membrane introduces a memory effect in the system causing, e.g., anomalous diffusion \citep{daddi16} or a sign reversal of two-particle hydrodynamic interactions \citep{daddi16c}.
On the experimental side, the near-wall mobility of a spherical particle has been investigated using optical tweezers \citep{faucheux94,lin00, dufresne01, schaffer07}, digital video microscopy \citep{eral10, cervantesMartinez11, dettmer14, traenkle16} and evanescent wave dynamic light scattering \citep{holmqvist07, michailidou09, lisicki12, rogers12, michailidou13, wang14, lisicki14}, where a significant alteration of particle motion has been observed in line with theoretical predictions.
The influence of a nearby elastic cell membrane has been further investigated using optical traps \citep{kress05, shlomovitz13, boatwright14, juenger15} and magnetic particle actuation \citep{irmscher12}.

Particles with a non-spherical shape, such as spheroids or rod-like particles, have also received researchers' attention.
The first attempt to investigate the Brownian motion of an anisotropic particle dates back to \cite{perrin34, perrin36} who computed analytically the drag coefficients for a spheroid diffusing in a bulk fluid.
A few decades later, \cite{batchelor70} pioneered the idea that the flow field surrounding a slender body, such as an elongated particle, may conveniently be represented by a line distribution of Stokeslets between the foci.
The method has successfully been applied to a wide range of external flows \citep{chwang75} and near boundaries such as a plane hard wall  \citep{demestre75, schiby80, mitchell15} or a fluid-fluid interface \citep{blake81}. Using the multipole expansion of the near-wall flow field, \cite{lisicki16} have shown that to leading order the mobility of an arbitrary axisymmetric particle near a hard wall can be expressed in closed form by combining the appropriate Green's function with the particle's bulk mobility. 
Direct-simulation numerical investigations of colloidal axisymmetric particles near a wall have been carried out using boundary integral methods \citep{hsu89}, stochastic rotation dynamics \citep{padding10, Neild_2010} and finite element methods \citep{decorato15}.

Diffusion of micrometer-sized ellipsoidal particles has been investigated experimentally using digital video microscopy \citep{han06, han09, zhongyu10, Neild_2010}.
Experiments on actin filaments have been conducted using fluorescence imaging and particle tracking \citep{li04} finding that the measured diffusion coefficients can appropriately be accounted for by a correction resting on the hydrodynamic theory of a long cylinder confined between two walls.
The  confined rotational diffusion coefficients of carbon nanotubes have been measured using fluorescence video microscopy \citep{duggal06} and optical microscopy \citep{bhaduri08}, where a reasonable agreement has been reported with theoretical predictions.
More recently, the three-dimensional rotational diffusion of nanorods \citep{cheong10} and rod-like colloids have been measured using video \citep{colin12} and confocal microscopy \citep{mukhija07}.

Yet, to the best of our knowledge, motion of a non-spherical particle in the vicinity of deformable elastic interfaces has not been studied so far.
In this contribution, we examine the dynamics of an axisymmetric particle near a red blood cell (RBC) membrane using theoretical predictions in close combination with fully resolved boundary integral simulations.
{The results of the present theory may be used in microrheology experiments in order to characterize the mechanical properties of the membrane \citep{waigh16}. }

The paper is organised as follows. 
In Sec.~\ref{model}, we formulate the theoretical framework for the description of the motion of a colloidal particle in the vicinity of an elastic membrane. We introduce the notion of hydrodynamic friction, mobility, and a model for the membrane. 
In Sec.~\ref{tensors} we outline the mathematical derivation of the correction to the bulk mobility tensor of the particle due to the presence of an interface and provide explicit expressions for the correction valid for any axially symmetric particle. 
In Sec.~\ref{bim}, we describe the boundary integral method (BIM) used to numerically compute the components of the mobility tensor. 
Sec.~\ref{results} contains a comparison of analytical predictions and numerical simulations for a spheroidal particle, followed by concluding remarks in Sec.~\ref{conclusions}. The mathematical details arising in the course of the work are discussed in the Appendices.

\section{Hydrodynamics near a membrane}\label{model}

\begin{figure}
 \begin{center}
  \scalebox{1.6}{\input{Pics/spheroidNearMembrane}}
 \caption{Illustration of a spheroidal particle located at $z = z_0$ above an elastic cell membrane. 
 The short and long axes are denoted by $a$ and $c$, respectively.
 The unit vector $\vect{u}_1$ is pointing along the spheroid symmetry axis and $\vect{u}_{2}$ is the unit vector perpendicular to the plane of the figure.
The unit vector $\vect{u}_{3}$ is defined to be orthogonal to both $\vect{u}_1$ and $\vect{u}_{2}$.}
 \label{particleIllustration}
 \end{center}
\end{figure}
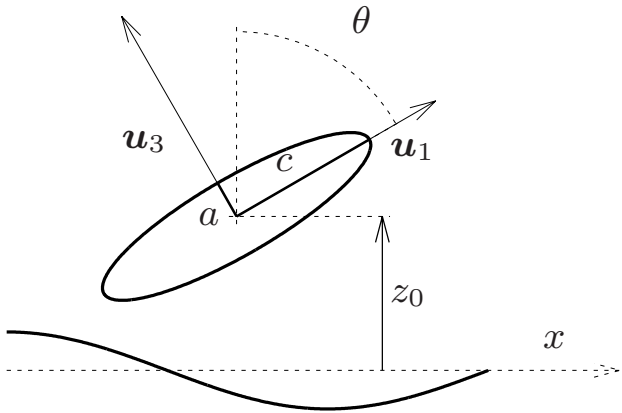

We consider an axially symmetric particle immersed in an incompressible Newtonian fluid, moving close to an elastic membrane. 
{The fluid is assumed to have the same dynamic viscosity $\eta$ on both sides of the membrane.}
As an example, we will focus later on a prolate spheroidal particle as shown in figure~\ref{particleIllustration}. 
The position of the centre of the particle is $\BR$, while its orientation is described by the unit vector $\uu_1$ pointing along the symmetry axis.
The laboratory frame is spanned by the basis vectors $\{\ex,\ey,\ez\}$. 

We denote by $z_0$ the vertical distance separating the centre of the particle from the undisplaced membrane located at the plane $z=0$ and extended infinitely in the horizontal plane~$xy$.
It is convenient to introduce the body-fixed frame of reference, formed by the three basis vectors {$\{\uu_1,\uu_{2},\uu_{3}\}$}. The unit vector $\uu_{2}$ is parallel to the undisplaced membrane and perpendicular to the particle axis, and $\uu_{3}$ completes the orthonormal basis. We define $\theta$ as the angle between $\uu_1$ and $\ez$ such that $\cos \theta =\ez\,\mathbf{\cdot}\,\uu_1$.
The basis vectors in the particle frame are then given by $\uu_{2}=(\ez\times \uu_1)/\left\vert \ez\times \uu_1 \right\vert$ and $\uu_{3}=\uu_1 \times \uu_{2}$.

In the inertia-free regime of motion, the fluid dynamics are governed by the stationary incompressible Stokes equations
\begin{align}
   \eta \boldNabla^2 \vect{v} (\R) - \boldNabla p (\R) + \vect{f} (\R) &= 0, \label{StokesEquations_1}
   \\ \boldNabla\,\mathbf{\cdot}\,\vect{v} (\R) &= 0 \, , \label{StokesEquations_2}
\end{align}
where $\vect{v}$ is the fluid velocity, $p$ is the pressure field and $\vect{f}$ is the force density acting on the fluid due to the presence of the particle. 
We omit the unsteady term in the Stokes equations, since in realistic situations it leads to a negligible contribution to the mobility corrections \citep{daddi16}. 
For a discussion accounting for the unsteady term in bulk flow, see recent work by \cite{felderhof13}.
The flow $\bv(\br)$ may be superposed with an arbitrary external flow $\bv_0(\br)$ being a solution to the homogeneous Stokes equations in the absence of the particle. 

Consider now a colloidal particle near the membrane. 
The total force $\BF$, torque $\BT$ and stresslet (symmetric force dipole) $\BS$ are linearly related to the velocities (translational $\BV$ and angular $\oOmega$) of the particle relative to an external flow by the generalised friction tensor \citep{kim13}
\begin{equation}\label{friction_single}
\begin{pmatrix}
\BF \\
\BT \\
\BS
\end{pmatrix} = 
\begin{pmatrix}
\bzz^{tt} & \bzz^{tr} & \bzz^{td} \\
\bzz^{rt} & \bzz^{rr} & \bzz^{rd} \\
\bzz^{dt} & \bzz^{dr} & \bzz^{dd} \\
\end{pmatrix}
\begin{pmatrix}
\bv_0-\BV  \\
 \bm{\omega}_0-\oOmega \\
\BE_0
\end{pmatrix},
\end{equation}
with $\bv_0=\bv_0(\BR)$, the vorticity $\bm{\omega}_0=\frac{1}{2}\boldNabla\times\bv_0(\BR)$, and the rate of strain $\BE_0=\overline{\bm{\nabla}\bv_0}(\BR)$ of the external flow (the bar denotes the symmetric and traceless part of the velocity gradient).

A complimentary relation defines the generalised mobility tensor
\begin{equation}\label{mobility_single}
\begin{pmatrix}
\BV-\bv_0 \\
 \oOmega-\bm{\omega}_0\\
-\BS
\end{pmatrix} =  
\begin{pmatrix}
\mi^{tt} & \mi^{tr} & \mi^{td} \\
\mi^{rt} & \mi^{rr} & \mi^{rd} \\
\mi^{dt} & \mi^{dr} & \mi^{dd} \\
\end{pmatrix}
\begin{pmatrix}
\BF  \\
 \BT \\
\BE_0
\end{pmatrix}.
\end{equation}
Upon examining Eqs. (\ref{friction_single}) and (\ref{mobility_single}), we note that the $6\times6$ mobility tensor $\mi$ is the inverse of the friction tensor $\bzz$
\begin{equation} \label{frictionmobility}
\bzz^{-1}=\begin{pmatrix}
\bzz^{tt} & \bzz^{tr} \\
\bzz^{rt} & \bzz^{rr} 
\end{pmatrix}^{-1} 
=
\begin{pmatrix}
\mi^{tt} & \mi^{tr} \\
\mi^{rt} & \mi^{rr} 
\end{pmatrix}= \mi \, .
\end{equation}
Relations between other elements of the generalised mobility and friction tensors may be found directly from Eqs. \eqref{friction_single} and \eqref{mobility_single}. These are general properties of Stokes flows following from the linearity of the governing equations. Finding an explicit form of these tensors requires the solution of Stokes equations \eqref{StokesEquations_1} and \eqref{StokesEquations_2} with appropriate boundary conditions on the confining interfaces. Since we aim at computing the particle mobility nearby a membrane endowed with surface elasticity and bending resistance, a relevant model for the membrane dynamics needs to be introduced at this point.

The Skalak model \citep{skalak73} is well-established and commonly used to represent RBC membranes \citep{krueger11, Freund_2014}. The elastic properties of the interface are characterised by two moduli: elastic shear modulus $\kS$ and area dilatation modulus $\kA$. Resistance towards bending has been further included following the model of \cite{helfrich73} with the associated bending modulus $\kB$. In this approach, the linearized tangential and normal traction jumps across the membrane are related to the membrane displacement field $\bs{h}$ at $z=0$ and the dilatation $\epsilon$ by \citep{daddi16}  
      \begin{align}
      [\sigma_{z\alpha}] &= -\frac{\kS}{3} \left( \DeltaPara h_\alpha + (1+2C) \partial_{\alpha} \epsilon \right) \,, \quad \alpha \in \{ x,y \} \, , \label{tangentialCondition}\\
      [\sigma_{zz}] &= \kB \DeltaPara^2 h_z \, , \label{normalCondition}
      \end{align}
where  $[f]:=f(z=0^{+})-f(z=0^{-})$ denotes the jump of the quantity $f$ across the membrane. The dilatation $\epsilon:=\partial_x h_{x}+\partial_y h_{y}$ is the trace of the strain tensor.
The Skalak parameter is defined as $C := \kA/\kS$.
Here $\DeltaPara := \partial^2_{x} + \partial^2_{y}$ is the Laplace-Beltrami operator along the membrane.
The components $\sigma_{z\alpha}$ of the stress tensor in the fluid are expressed in a standard way by
 $\sigma_{z\alpha} = -p \delta_{z\alpha} + \eta (\partial_\alpha v_{z} + \partial_z v_{\alpha})$ for  $\alpha\in \{x,y,z\}$ \citep{kim13}. 

{ 
The membrane displacement $\vect{h}$ and the fluid velocity $\vect{v}$ are related by the no-slip boundary condition at the undisplaced membrane, which in Fourier space takes the form 
\begin{equation}
 {v}_\alpha = i\omega {h}_\alpha |_{z=0} \, , \quad \alpha \in \{ x,y,z \} \, ,
\end{equation}}
with $\omega$ being the characteristic frequency of forcing in the system. 
The frequency-dependent elastic deformation effects are characterised by two dimensionless parameters, as described in \cite{daddi16}
\begin{equation}
\beta = \frac{12z_0 \eta \omega}{\kS+\kA}, \qquad\qquad \betaB = 2z_0 \left(\frac{4\eta\omega}{\kB}\right)^{1/3}, \label{beta_and_betaB_def}
\end{equation}
{Further details of the derivation of $\beta$ and $\betaB$ can be found in Appendix~\ref{greens}.}
The effect of shear resistance and area dilatation is thus captured by $\beta$, while $\betaB$ describes the bending resistance of the membrane. In the steady case for which  $\beta=\betaB=0$, corresponding to a vanishing frequency or to an infinitely stiff membrane, we expect to recover the results for a hard no-slip wall.

In the case of periodic forcing or time-dependent deformation of the membrane, the quantity of interest is the frequency-dependent mobility tensor. Our aim in this work is to find all the components of $\mi(\omega)$ for an axisymmetric particle close to an elastic membrane. Accordingly, due to the presence of the interface, the near-membrane mobility will then have a correction on top of the bulk mobility $\mi_0$, 
\begin{equation}
\mi(\omega) = \mi_0 + \bm{\Delta}\mi(\omega) \, , 
\end{equation}
stemming from the interaction of the flow created by the particle with the boundary. To determine the form of $\mi(\omega)$ in an approximate manner, we use the results by \cite{lisicki16} valid for a hard no-slip wall and generalize them to the case of an elastic membrane. 
Their idea is based on a multipole expansion \citep{cichocki00} of the flow field around an axially symmetric particle close to a boundary, with {a corresponding expansion} of the force distribution on its surface. If the particle is sufficiently far away from the wall, they have shown that the dominant correction to its friction matrix can be viewed as an interaction between the centre of the particle and its hydrodynamic image. They provide explicit expressions for the elements of the friction tensor for all types of motion (translation, rotation and coupling terms) which yields the corrected mobility tensor upon inversion. The same route may be followed for a membrane, provided that the form of the Green's tensor for the system is known.

A general Stokes flow can be constructed using the Green's function $\Gmatr(\br,\br')$ being the solution of Eqs.~\eqref{StokesEquations_1} and \eqref{StokesEquations_2} subject to a time-dependent point force $\bs{f}(r)= \bs{F}(t)\delta(\bs{r}-\bs{r}' )$ with the appropriate boundary conditions on the membrane. 
In an unbounded fluid, the Green's tensor is the Oseen tensor \citep{kim13} $\Gmatr_0(\br,\br')=\Gmatr_0(\br-\br')$, with $\Gmatr_0(\br) = \left(\bm{1} + {\br}{\br}/r^2\right)/(8\upi\eta r)$, with $r:=|\br|$. In the presence of boundaries, the Green's tensor contains the extra term $\bs{\Delta}\Gmatr$ describing the flow reflected from the membrane, so that $\Gmatr= \Gmatr_0 + \bs{\Delta}\Gmatr$.

The exact Green's function for a point force close to a membrane has recently been computed by some of us in \citep{daddi16}.
For the resolution of Eqs.~\eqref{StokesEquations_1} and \eqref{StokesEquations_2} with a point force acting at $\BR = (0,0,z_0)$, the two-dimensional Fourier transform in the $xy$ plane was used to solve the resulting equations with accordingly transformed boundary conditions. 
The procedure has been previously described in detail and therefore we only list {the main steps for the determination of} the Green's tensor in Appendix~\ref{greens} of this work.

\section{Near-membrane mobility tensors}\label{tensors}

We search for the near-membrane mobility tensor, $\mi(\omega) = \mi_0 + \bm{\Delta}\mi(\omega)$ by calculating the leading-order correction to the bulk mobility. 
To this end, we follow the route outlined in a recent contribution by \cite{lisicki16} who derived analytic expressions for the friction tensor of an axially symmetric particle in the presence of a hard no-slip wall. 
The friction tensor, similarly to the mobility tensor, can be split into the bulk and the correction term
\begin{equation}
\bzz=\bzz_0+\bs{\Delta}\bzz.
\end{equation} 
The final expressions for the corrected friction tensor involve elements of the bulk friction tensor of the particle, and the distance- and orientation-dependent (derivatives of) the appropriate Green's function.
For the hard no-slip wall treated in \cite{lisicki16} the latter is the Blake tensor \citep{blake71} while in the present case the frequency-dependent Green's functions from \cite{daddi16} are employed.
The expressions for the friction tensor with a general Green's function read \citep{lisicki16}
\begin{align}\label{corrtt} 
\bm{\Delta}\bzz^{tt} =& -\frac{1}{8\upi\eta}\frac{1}{2z_0} \bzz^{tt}_0\bg^{tt}\bzz^{tt}_0 + \frac{1}{(8\upi\eta)^2}\frac{1}{(2z_0)^2} \bzz^{tt}_0\bg^{tt}\bzz^{tt}_0\bg^{tt}\bzz^{tt}_0 +\textit{O}(z_0^{-3}), \\
\bm{\Delta}\bzz^{tr} =& -\frac{1}{8\upi\eta}\frac{1}{(2z_0)^2} \bzz^{tt}_0\bg^{td}\bzz^{dr}_0 + \textit{O}(z_0^{-3}), \\
\bm{\Delta}\bzz^{rt} =& -\frac{1}{8\upi\eta}\frac{1}{(2z_0)^2} \bzz^{rd}_0\bg^{dt}\bzz^{tt}_0 + \textit{O}(z_0^{-3}), \\ \label{corrrr}
\bm{\Delta}\bzz^{rr} =& -\frac{1}{8\upi\eta}\frac{1}{(2z_0)^3}[\bzz^{rr}_0\bg^{rr}\bzz^{rr}_0 + \bzz^{rr}_0\bg^{rd}\bzz^{dr}_0+ \bzz^{rd}_0\bg^{dr}\bzz^{rr}_0+\bzz^{rd}_0\bg^{dd}\bzz^{dr}_0] + \textit{O}(z_0^{-4}).
\end{align}
where the directional tensors $\bg$ are defined by
\begin{equation}
\bs{\Delta}\Gmatr^{\gamma\delta} = \frac{1}{8\upi\eta}\frac{1}{(2z_0)^{a}} \bg^{\gamma\delta}.
\end{equation}
Here, $\bs{\Delta}\Gmatr^{\gamma\delta}$ are the multipole elements of the Green's integral operator which will be derived below. Further, $\gamma,\delta\in\{t,r,d\}$ and $a=1$ for $tt$, $a=2$ for $(tr,rt,td,dt)$ and $a=3$ for $(dr,rd,rr,dd)$ parts. 
In Eqs.~\eqref{corrtt}-\eqref{corrrr} it should be understood that the tensors are appropriately contracted to yield second-order tensor corrections.

We now apply this result to our system. Our goal is to obtain explicit expressions for the mobility tensors for an axially symmetric particle in the presence of a membrane in terms of its bulk mobilities.
This can be done in two steps.

Firstly, we invert the friction relations \eqref{corrtt}-\eqref{corrrr}, as detailed in Appendix~\ref{appInversion}, to obtain  analogous relations for the mobilities:
\begin{align}\label{Mtt}
\bm{\Delta}\mi^{tt} &= \frac{1}{8\upi\eta} \frac{1}{2z_0}\bg^{tt} + \textit{O}(z_0^{-3}), \\ \label{Mtr}
\bm{\Delta}\mi^{tr} &= - \frac{1}{8\upi\eta} \frac{1}{(2z_0)^2}\bg^{td}\mi_0^{dr} + \textit{O}(z_0^{-3}), \\ \label{Mrt}
\bm{\Delta}\mi^{rt} &=  \frac{1}{8\upi\eta} \frac{1}{(2z_0)^2} \mi_0^{rd}\bg^{dt} + \textit{O}(z_0^{-3}), \\ \label{Mrr}
\bm{\Delta}\mi^{rr} &= \frac{1}{8\upi\eta} \frac{1}{(2z_0)^3}\big[\bg^{rr} - \mi_0^{rd}\bg^{dr}  + \bg^{rd}\mi_0^{dr}
                     - \mi_0^{rd}\bg^{dd}\mi_0^{dr} \big] + \textit{O}(z_0^{-4}).
\end{align}

These expressions allow straightforward computation of the near-membrane mobilities for arbitrarily-shaped axisymmetric particles if their bulk mobilities are known, either numerically or analytically. 
Compared to a numerical inversion of the friction tensor, which in principle would be preferable as it avoids the possibility of negative mobilities \citep{lisicki16}, this approach has the advantage that explicit analytical expressions for the mobility can be obtained.

Remarkably, the final formulae include only one bulk characteristic of the particle, namely the tensors $\mi_0^{rd}$ and $\mi_0^{dr}$  which describe the rotational motion of the particle in response to elongational flow. This form follows from the particular symmetries of an axially symmetric particle with inversional symmetry ($\uu_1 \leftrightarrow-\uu_1$). 

Secondly, to obtain the directional tensors $\bg$, we consider a general Green's tensor $\Gmatr(\br,\br')= \Gmatr_0(\br-\br') + \bs{\Delta}\Gmatr(\br,\br')$ and a body placed at $\BR$ with a force distribution $\vect{f}(\br)$ on its surface. The flow at a point $\br$ due to this forcing may be written as the integral equation 
\begin{equation}\label{green}
\bv(\br) = \integr{}{}{\br'} \Gmatr(\br,\br')\mathbf{\cdot}\vect{f}(\br') \,
\end{equation}
with the integral performed over the surface of the body. 
The idea of the derivation of the correction is to find, given the force density, the flow incident on the particle itself due to the presence of an interface. 
Thus we consider Eq. \eqref{green} with only the membrane-interaction part $\bs{\Delta}\Gmatr(\br,\br')$ of the Green's tensor and expand it in both arguments around $\br=\br'=\BR$. 
The integrals of the subsequent terms on the RHS reproduce the force multipole moments, while the expansion of the LHS yields the multipole expansion of the flow field. By matching the relevant multipoles, we find explicit expressions for the $\bs{\Delta}\Gmatr^{\gamma\delta}$, with $\gamma,\delta\in\{t,r,d\}$, as described in \cite{lisicki16}. The resulting formulae are
 \begin{align}
\Delta\mathcal{G}^{tt}_{\alpha\beta} & = \lim_{\br,\br'\to\BR} \Delta\mathcal{G} _{\alpha\beta}\, , \label{DeltaGtt_ab} \\
\Delta\mathcal{G}^{tr}_{\alpha\beta} & =\lim_{\br,\br'\to\BR} \frac{1}{2} \epsilon_{\alpha\mu\nu} \partial_\mu  \Delta\mathcal{G}_{\nu\beta}\, , \\
\Delta\mathcal{G}^{rt}_{\alpha\beta} & = \lim_{\br,\br'\to\BR}-\frac{1}{2} \epsilon_{\mu\nu\beta} \partial'_\nu \Delta\mathcal{G}_{\alpha\mu}\, ,  \\
\Delta\mathcal{G}^{rr}_{\alpha\beta} & = \lim_{\br,\br'\to\BR}\frac{1}{4} \epsilon_{\alpha\mu\gamma}\epsilon_{\beta\nu\eta} \partial_\gamma \partial'_\eta \Delta\mathcal{G}_{\mu\nu}\, , \\
\Delta\mathcal{G}^{td}_{\alpha\beta\gamma} & =\lim_{\br,\br'\to\BR} \partial'_\gamma \Delta\mathcal{G}_{\alpha\beta}\, ,  \\
\Delta\mathcal{G}^{dt}_{\alpha\beta\gamma} & =\lim_{\br,\br'\to\BR} \partial_\alpha \Delta\mathcal{G}_{\beta\gamma}\, ,  \\
\Delta\mathcal{G}^{dr}_{\alpha\beta\gamma} & = \lim_{\br,\br'\to\BR}-\frac{1}{2}\epsilon_{\gamma \mu \nu}\partial'_\nu {\overbracket{\partial_\alpha \Delta\mathcal{G}_{\beta \mu}}}^{(\alpha\beta)} \, ,  \\                       
\Delta\mathcal{G}^{rd}_{\alpha\beta\gamma} & =\lim_{\br,\br'\to\BR} \frac{1}{2}\epsilon_{\alpha \mu \nu}\partial'_\mu {\overbracket{\partial_\beta \Delta\mathcal{G}_{\nu\gamma}}}^{(\beta\gamma)} \, ,  \\                          
\Delta\mathcal{G}^{dd}_{\alpha\beta\gamma\delta} &=\lim_{\br,\br'\to\BR} {\overbracket{\partial_\alpha\partial'_\delta \Delta\mathcal{G}_{\beta\gamma}}}^{(\alpha\beta)(\gamma\delta)}\, , \label{DeltaGdd_abcd}
\end{align}
where $\epsilon_{\alpha\mu\nu}$ is the Levi-Civita tensor and the symbol ${\overbracket{\phantom{...}}}^{(\alpha\beta)}$ denotes the symmetric and traceless part with respect to indices $\alpha,\beta$. Explicitly, the reductions for an arbitrary 3rd and 4th order traceless tensor read
\begin{align}\label{reductions}
{\overbracket{M_{\alpha\beta\gamma}}}^{(\alpha\beta)}&=\frac{1}{2}\left( M_{\alpha\beta\gamma} + M_{\beta\alpha\gamma} \right), \nonumber \\
{\overbracket{M_{\alpha\beta\gamma\delta}}}^{(\alpha\beta)(\gamma\delta)}&=\frac{1}{4}\left( M_{\alpha\beta\gamma\delta} + M_{\beta\alpha\gamma\delta} + M_{\alpha\beta\delta\gamma} + M_{\beta\alpha\delta\gamma} \right) \, .\nonumber
\end{align}
The prime denotes a derivative with respect to the second argument. 
We note that the tensors $\bs{\Delta}\Gmatr^{dr,rd,dd}$ are traceless due to the incompressibility of the fluid, and therefore the trace needs not be subtracted in the procedure of symmetrization.
We further remark that Eqs.~\eqref{DeltaGtt_ab} through \eqref{DeltaGdd_abcd} involve differentiations and elementary operations that are well defined for complex quantities, and hence lead to convergent limits.

It is most natural to consider the correction in the reference frame of the particle, spanned by the three unit basis vectors {$\{\uu_1,\uu_{2},\uu_{3}\}$}.
In this frame, the mobility tensors of an axisymmetric particle have the form
\begin{equation}
\label{mutt}
\bs{\Delta}\mi^{tt,rr}=%
\begin{pmatrix}
\Delta\mu^{tt,rr}_{11} & 0 & \Delta\mu^{tt,rr}_{13} \\ 
0 &  \Delta\mu^{tt,rr}_{22} & 0 \\ 
\Delta\mu^{tt,rr}_{13} & 0 & \Delta\mu^{tt,rr}_{33}
\end{pmatrix} 
\end{equation} 
for translational and rotational motion, while the translation-rotation coupling tensor reads
\begin{equation} \label{mutr}
\bs{\Delta}\mi^{tr}=%
\begin{pmatrix}
0 & \Delta \mu^{tr}_{12} & 0 \\ 
0 & 0 & \Delta\mu^{tr}_{23} \\ 
0 & \Delta\mu^{tr}_{32} & 0%
\end{pmatrix}.
\end{equation}
The rotation-translation coupling tensor $\bs{\Delta}\mi^{rt}$ is obtained by simply taking the transpose of the translation-rotation coupling tensor given above. (See Supplemental Material at [URL will be inserted by publisher] for the frequency-dependent mobility corrections expressed in LAB frame).

\section{Boundary Integral Method}\label{bim}

Here we introduce the boundary-integral method \citep{pozrikidis01} used to numerically compute the mobility tensor of a truly extended spheroidal particle. 
The method is perfectly suited for treating 3D problems with complex, deforming boundaries such as RBC membranes in the Stokes regime \citep{zhao11, zhu14}.
In order to solve for the particle motion, given an applied force or torque, we combine a completed double layer boundary integral method (CDLBIM) \citep{power87} to the classical BIM \citep{zhao12}. 
The resulting equations are then discretised and transformed into a system of algebraic equations as detailed in~\citep{daddi16b, guckenberger16}.

For the numerical determination of the particle mobility components, a harmonic force $\vect{F}(t) = \vect{A} \ue^{\ui\omega_0 t}$ or torque $\vect{T}(t) = \vect{B} \ue^{\ui\omega_0 t}$ is applied at the particle surface.
After a short transient evolution, the particle linear and angular velocities can be described as $\vect{V} (t) = \vect{C} \ue^{\ui(\omega_0 t + \delta_t)}$ and $\boldsymbol{\Omega} (t) = \vect{D} \ue^{\ui(\omega_0 t + \delta_r)}$ respectively.
The amplitudes and phase shifts can be determined accurately by fitting the numerically recorded velocities using the trust region method \citep{conn00}.
In the LAB frame, the components $\mu_{\alpha\beta}^{tt}$ and $\mu_{\alpha\beta}^{rt}$ of the mobility are determined for a torque-free particle as
\begin{equation}
 \mu_{\alpha\beta}^{tt} = \frac{C_\alpha}{A_\beta} \ue^{\ui\delta_t}  \, , \qquad\qquad \mu_{\alpha\beta}^{rt} = \frac{D_\alpha}{A_\beta} \ue^{\ui\delta_r} \, .
\end{equation}
For a force-free particle, the components $\mu_{\alpha\beta}^{tr}$ and $\mu_{\alpha\beta}^{rr}$ are obtained from
\begin{equation}
 \mu_{\alpha\beta}^{tr} = \frac{C_\alpha}{B_\beta} \ue^{\ui\delta_t}  \, , \quad \mu_{\alpha\beta}^{rr} = \frac{D_\alpha}{B_\beta} \ue^{\ui\delta_r} \, .
\end{equation}

\section{Spheroid close to a membrane: theoretical and numerical results}\label{results}

In this Section, we present a comparison of our theoretical results to numerical simulations using the example of a prolate spheroidal particle. 
To begin with, we discuss the bulk mobility of a spheroid. 
Further on, we show the explicit form of the correction, and finally compare the components of the corrected mobility matrix to numerical simulations.

The bulk translational and rotational mobility tensors of a general axisymmetric particle have the form
\begin{equation}\label{bulk_mttrr}
\mi_0^{tt,rr} = \mu_\parallel^{t,r} \uu_1 \uu_1 + \mu_\perp^{t,r} (\bm{1} - \uu_1 \uu_1).
\end{equation}
The third-order tensors $\mi_0^{rd}$ and $\mi_0^{dr}$ have the Cartesian components
\begin{align}
(\mu_0^{rd})_{\alpha\beta\gamma} &= {\mu^{rd}} u_\sigma {\overbracket{\epsilon_{\sigma\alpha\beta} u_\gamma}}^{(\beta\gamma)} \, , \\
(\mu_0^{dr})_{\alpha\beta\gamma} &=  
{\mu^{dr}}  {\overbracket{u_\alpha \epsilon_{\beta\gamma\sigma} }}^{(\alpha\beta)}  u_\sigma \, , 
\end{align}
where, following from the Lorentz reciprocal theorem \citep{kim13}
\begin{equation}
\mu^{dr} = \mu^{rd} =: \lambda \, .
\end{equation} 
Note that due to the axial and inversional symmetry in bulk, we have $\mi_0^{tr}=\mi_0^{rt}=0$ and $\mi_0^{td}=\mi_0^{dt}=0$.

For a prolate spheroidal particle of eccentricity $e$, 
analytical results are available and the bulk mobility coefficients are given by (\cite{kim13}, table 3.4) 
 \begin{align}
  \mu_\parallel^{t} &= \frac{1}{6\upi\eta c} \frac{3}{8} \frac{-2e + (1+e^2)L}{e^3} \, , \\
  \mu_\perp^{t} &= \frac{1}{6\upi\eta c} \frac{3}{16} \frac{2e + (3e^2 - 1)L}{e^3} \, , \\
   \mu_\parallel^{r} &= \frac{1}{8\upi\eta c^3} \frac{3}{4} \frac{2e-(1-e^2)L}{e^3 (1-e^2)} \, , \\
  \mu_\perp^{r} &= \frac{1}{8\upi\eta c^3} \frac{3}{4} \frac{-2e+(1+e^2)L}{e^3 (2-e^2)} \, , 
 \end{align}
where $a$ and $c$ are the short and long axis of the spheroid and
\begin{equation}
 e = \sqrt{1-\left(\frac{a}{c}\right)^2} \, , \quad L = \ln \left( \frac{1+e}{1-e} \right).
\end{equation}
To obtain the final ingredient $\mu^{rd}$, we observe from the definitions in Eqs. \eqref{friction_single} and \eqref{mobility_single} that $\mu_{\alpha\beta\gamma}^{rd} = \mu_{\alpha\delta}^{rr} \zeta_{\delta\beta\gamma}^{rd}$ and $\mu_{\alpha\beta\gamma}^{dr} = - \zeta_{\alpha\beta\delta}^{dr} \mu_{\delta\gamma}^{rr}$, leading to 
\begin{equation}
\mu^{rd}  = \mu_\perp^{r} \zeta^{rd}.
\end{equation}
The component $rd$ of the friction tensor is \citep{kim13}
\begin{equation}
 \frac{ \zeta^{rd}}{8\upi\eta c^3} =  \frac{4}{3} \frac{e^5}{-2e + (1+e^2)L} \,.
\end{equation}
Therefore we obtain the $rd$ coefficient of the mobility tensor 
\begin{equation}
 \lambda = \frac{e^2}{2-e^2} \, .
\end{equation}

\begin{figure*}
 \begin{center}
  \scalebox{0.7}{\input{Pics/DeltaMu_Tran}}
  \caption{(Color online) The scaled translational mobility correction components versus the scaled frequency. 
 The spheroid is located above the membrane at $z_0 = 2c$ inclined at an angle $\theta=\upi/3$ from the vertical.
 The analytical  predictions of the real and imaginary parts of the translational mobility corrections are shown as dashed and solid lines, respectively.
 {The corrections due to shearing and bending are shown respectively in green (bright grey in a black and white printout) and red (dark grey in a black and white printout).}
 Horizontal dotted lines represent the hard-wall limits from \cite{lisicki16}. 
 BIM simulations are marked as squares and circles for the real and imaginary parts, respectively.
 {For the membrane parameters we take a reduced bending modulus $E_\mathrm{B} := \kB/(c^2\kS) = 2/3$ and the Skalak parameter $C=1$.} }
\label{DeltaMu_Tran}
\end{center}
\end{figure*}
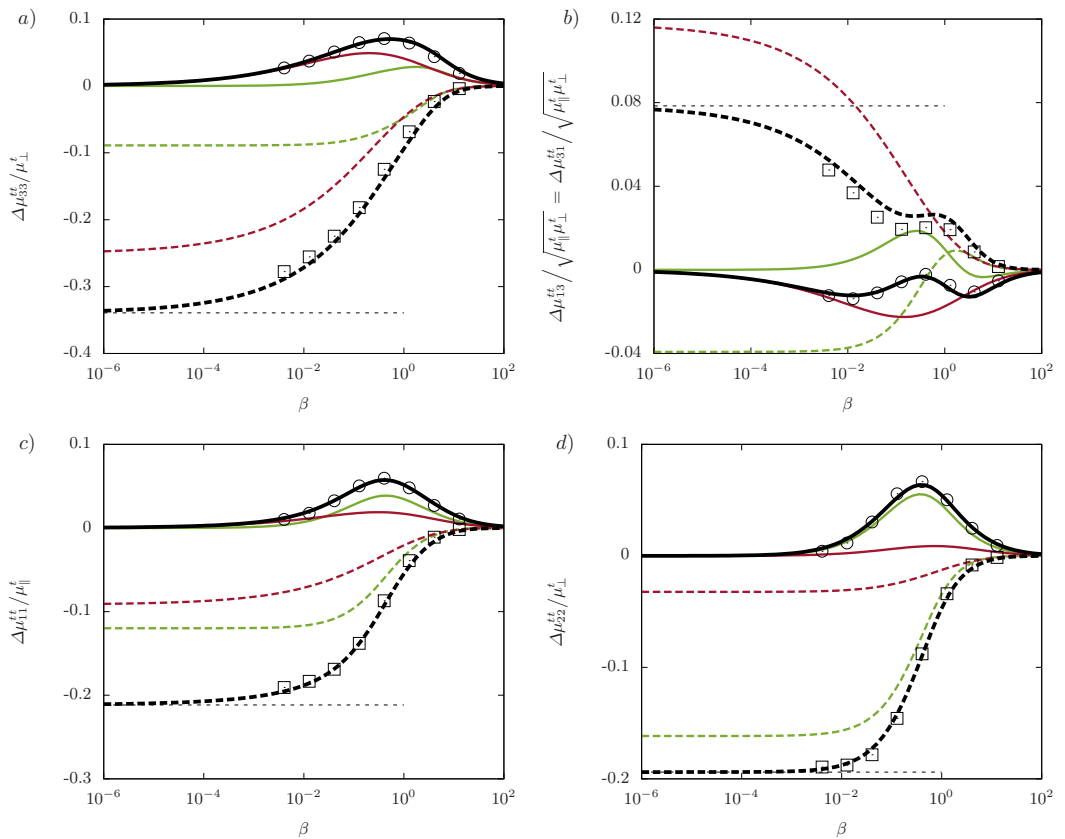

\begin{figure}
 \begin{center}
  \scalebox{0.7}{\input{Pics/DeltaMu_Coup}}
 \caption{(Color online) The scaled coupling mobility corrections versus the scaled frequency.
 Black and blue symbols refer to the $tr$ and $rt$ components, respectively, obtained from BIM simulations.
 The other colors are the same as in figure~\ref{DeltaMu_Tran}.}
 \label{DeltaMu_Coup}
 \end{center}
\end{figure}
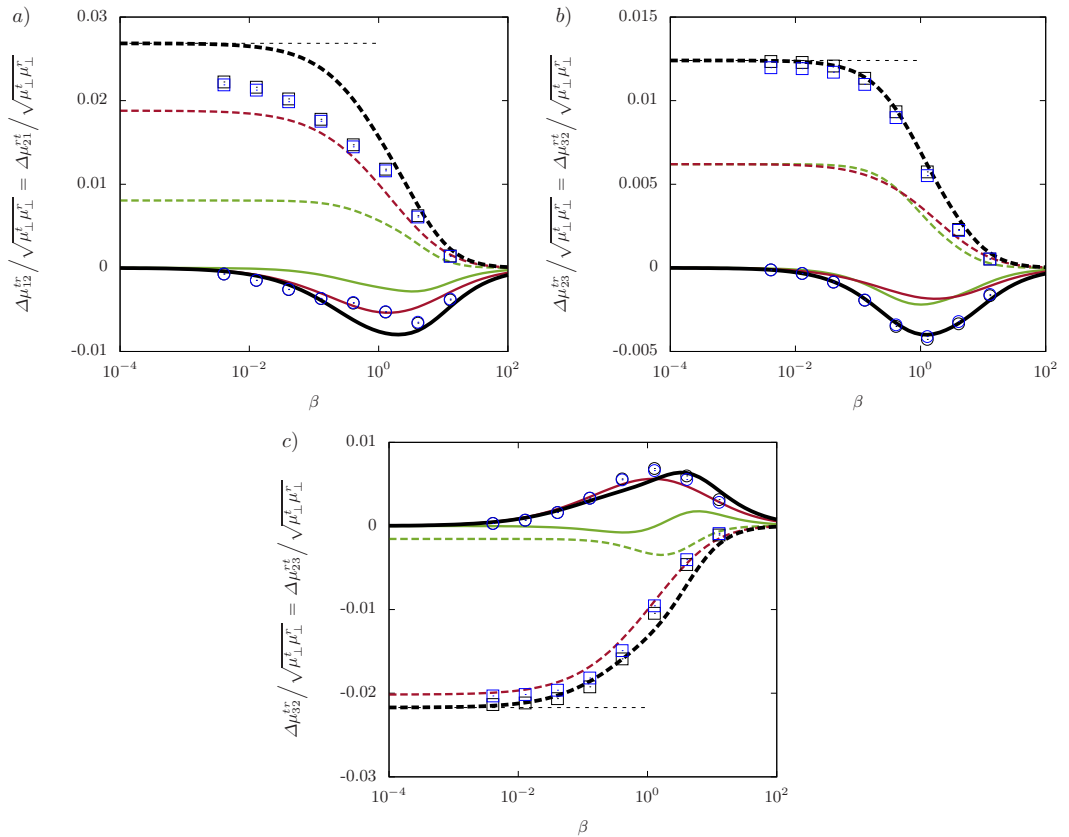

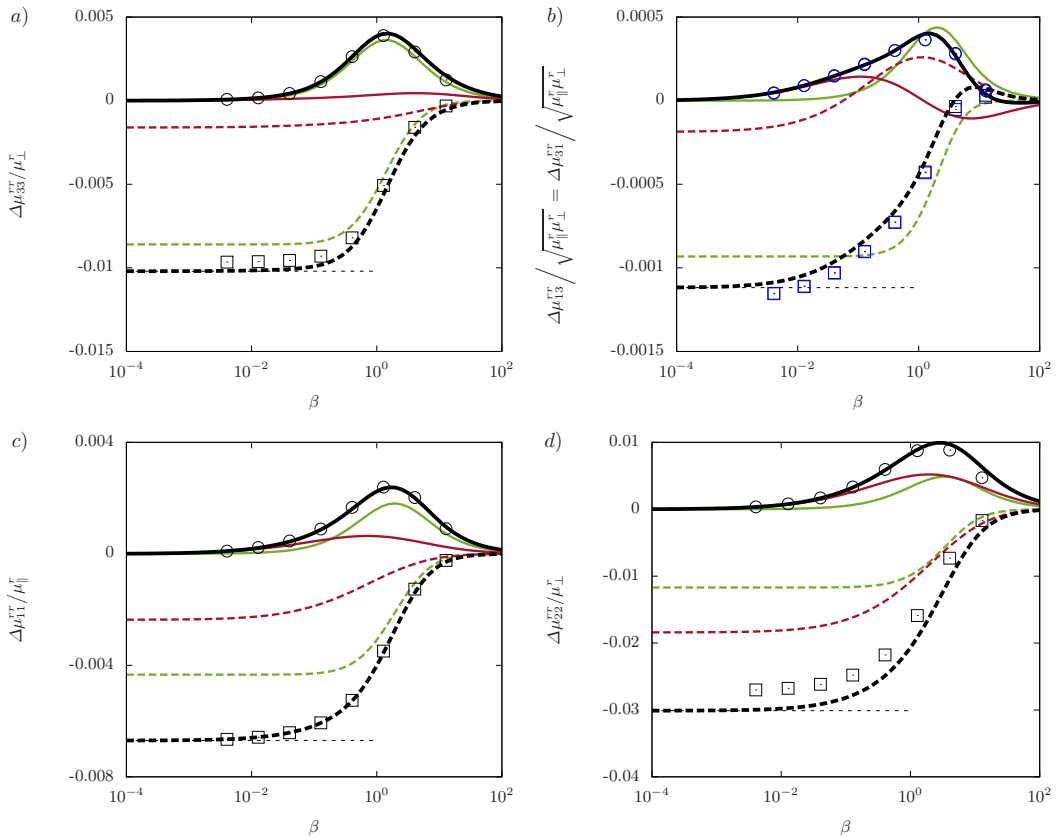
\begin{figure*}
 \begin{center}
  \scalebox{0.7}{\input{Pics/DeltaMu_Rot}}
 \caption{(Color online) The scaled rotational mobility correction component versus the scaled frequency.
 Black and blue symbols refer to the 13 and 31 components, respectively, obtained from BIM simulations.
 The color code is the same as in figure~\ref{DeltaMu_Tran}.}
 \label{DeltaMu_Rot}
 \end{center}
\end{figure*}

Having introduced the bulk hydrodynamic mobilities of a spheroid, we turn our attention to the membrane correction which in the frame of the particle can be written as in Eqs~\eqref{mutt} and~\eqref{mutr}.
We find that the corrections to the translational mobilities as given in general form in Eq.~\eqref{Mtt} can, for a spheroid, be written in closed form as
\begin{align}
8\upi\eta (2z_0) \Delta \mu_{11}^{tt} &=P\sin^2\theta + Q\cos^2\theta,  \label{deltaMu_tt_11} \\
8\upi\eta (2z_0) \Delta \mu_{13}^{tt} &=(P-Q)\sin\theta \cos \theta, \\
8\upi\eta (2z_0) \Delta \mu_{22}^{tt} &=P, \\
8\upi\eta (2z_0) \Delta \mu_{33}^{tt} &=P\cos^2\theta + Q\sin^2\theta, \label{deltaMu_tt_33}
\end{align}
 \label{translation-translation}
and $\Delta\mu_{13}^{tt}= \Delta\mu_{31}^{tt}$. Thus they have the desired symmetry of Eq. \eqref{mutt}. 
Expressions for $P(\beta,\betaB)=P_\mathrm{S}(\beta)+P_\mathrm{B}(\betaB)$ and $Q(\beta,\betaB)=Q_\mathrm{S}(\beta)+Q_\mathrm{B}(\betaB)$ are provided explicitly in Appendix \ref{analyticalExpressions}.

For the translation-rotation coupling, the non-vanishing mobility corrections as given by Eq.~\eqref{Mtr}, can be cast in the frame of the particle as
 \begin{align}
  8\upi\eta (2z_0)^2 \Delta \mu_{12}^{tr} &= \lambda \sin\theta \left( M + N  \cos^2 \theta \right) \, , \label{deltaMu_tr_12} \\
  8\upi\eta (2z_0)^2 \Delta \mu_{23}^{tr} &= \lambda M  \cos\theta \, , \label{deltaMu_tr_23} \\
  8\upi\eta (2z_0)^2 \Delta \mu_{32}^{tr} &= -\lambda\cos\theta \left( M  + N  \sin^2 \theta \right) \, , \label{deltaMu_tr_32}
 \end{align}
 \label{translation-rotation-coupling}
where $M$ and $N$ are now functions of the parameters $\beta$ and $\betaB$, and can likewise be decomposed into shearing and bending contributions. The dependence on the bulk $rd$ mobility $\lambda$ is explicitly separated out.

Finally, considering the rotational part as stated by Eq.~\eqref{Mrr},
the non-vanishing components of the mobility correction in the frame of the particle can conveniently be cast in the following forms
 \begin{align}
  8\upi\eta (2z_0)^3 \Delta \mu_{11}^{rr} &= A_0  + A_2  \cos^2 \theta \, , \label{deltaMu_rr_11} \\
  8\upi\eta (2z_0)^3 \Delta \mu_{13}^{rr} &= D \sin\theta \cos \theta \, , \label{deltaMu_rr_13} \\
  8\upi\eta (2z_0)^3 \Delta \mu_{22}^{rr} &= C_0 + \lambda C_2  \cos^2\theta + \lambda^2 C_4  \cos^4 \theta \, , \label{deltaMu_rr_22} \\  
  8\upi\eta (2z_0)^3 \Delta \mu_{33}^{rr} &= H_0  + H_2 \cos^2 \theta ,  \label{deltaMu_rr_33}
 \end{align}
 \label{rotation-rotation}
and with $\Delta \mu_{13}^{rr} =\Delta \mu_{31}^{rr}$. 
All the functions depend on $(\beta,\betaB)$ and are decomposed into bending and shearing parts in appendix \ref{analyticalExpressions}. 
In addition, the functions $C$, $D$ and $H$ depend on the coefficient $\lambda$.

{It can be seen that the mobility corrections for an axisymmetric particle in their dominant terms possess a simple angular structure.
The latter stems from the contraction of the particle friction tensors (which have an axial symmetry, dictated by their shape, with respect to the body axis) with the vertical multipole components of the Blake tensor (which have the same structure but with respect to a different axis, i.e. the vertical direction). 
This contraction requires transformation of corresponding tensors into the common frame of reference, which generates simple polynomials in sine and cosine functions of the inclination angle as discussed in \cite{lisicki16}.}

In the following, we shall present a comparison between these analytical predictions and  numerical simulations using the Boundary Integral Method, presented in Sec.~\ref{bim}. 
We consider a prolate spheroid of aspect ratio $p := c/a = 2$, inclined at an angle $\theta=\upi/3$ to the $z$ axis, positioned at $z_0 = 2c$ above a planar elastic membrane. 
{For the membrane, we take a reduced bending modulus $E_\mathrm{B} := c^2\kS/\kB = 3/2$} for which the characteristic frequencies $\beta$ and $\betaB^3$ have the same order of magnitude. 
The Skalak parameter is $C=1$.
{Corresponding data showing the effect of the inclination angle and the reduced bending modulus can be found in the Supporting Information.}
{Our analytical predictions are applicable for large and moderate membrane-particle distances for which $c/z_0 \sim O(1)$ where we find good agreement with numerical simulations.}

Henceforth, the mobility corrections will be scaled by the associated bulk values. 
For diagonal terms, we choose the corresponding diagonal elements, namely $\mu^{t,r}_\parallel$ for $\mu^{t,r}_{11}$ and $\mu^{t,r}_\perp$ for $\mu^{t,r}_{22,33}$. 
For  non-diagonal terms, we use an appropriate combination of bulk mobilities, that is $\sqrt{\mu^{t,r}_{\parallel}\mu^{t,r}_{\perp}}$ for translations and rotations. 
The translation-rotation coupling tensors are scaled by $\sqrt{\mu^{t}_{\perp}\mu^{r}_{\perp}}$.

In figure~\ref{DeltaMu_Tran} we compare the components of the translational mobility  calculated from Eq.~\eqref{deltaMu_tt_11}-\eqref{deltaMu_tt_33} with those obtained from BIM simulations.
For the diagonal components we observe that the real part of the complex mobility corrections is monotonically increasing with frequency.
The imaginary part exhibits a non-monotonic bell-shaped dependence on frequency that peaks around $\beta\sim 1$.
The off-diagonal components 13 and 31 show a more complex dependence on frequency.
In the vanishing frequency limit, we recover the corrections near a hard-wall with stick boundary conditions recently calculated by  \cite{lisicki16}.
We further remark that for the present inclination of $\theta=\upi/3$ the components 33 and 13 are principally determined by bending resistance whereas shearing effect is more pronounced in the components 11 and 22.
A very good agreement is obtained between analytical predictions and numerical simulations for all components over the entire range of frequencies.

By examining the off-diagonal component 31 shown in figure~\ref{DeltaMu_Tran}~$b)$, it is clear that the shearing- and bending-related parts may have opposite contributions to the total translational mobility.
This observed trend implies that upon exerting a force along $\vect{u}_1$, there exists a drift motion along $\vect{u}_{3}$, either away or towards the membrane, depending on the shearing and bending properties.
In fact, for a membrane with bending-only resistance, such as a fluid vesicle, the spheroid is pushed away from the membrane in the same way as near a hard-wall.
On the other hand, for a membrane with shearing-only resistance, such as an artificial capsule, the motion is directed towards the membrane.

Figure~\ref{DeltaMu_Coup} shows the corrections to the translation-rotation coupling mobility versus the scaled frequency computed from Eq.~\eqref{deltaMu_tr_12} and \eqref{deltaMu_tr_32}. 
We observe that bending resistance is essentially the dominant contributor to the coupling mobility.
It can be shown that this trend is always the case regardless of spheroid orientation.
The BIM simulation results are consistent with the fact that the $tr$ and $rt$ mobility tensors are the transpose of each other and a good agreement is obtained between analytical predictions and simulations. 
The coupling terms are generally very small compared to the relevant bulk quantities.
This makes them somewhat more difficult to obtain precisely from the simulations which explains the small discrepancy notable in figure~\ref{DeltaMu_Coup}~a).


In figure~\ref{DeltaMu_Rot} we present the corrections to the components of the rotational mobility tensor as calculated by Eq.~\eqref{deltaMu_rr_11}-\eqref{deltaMu_rr_33} compared to the BIM simulations.
We remark that the shearing contribution manifests itself in a more pronounced way for the rotational mobilities. 
Moreover, the correction to the rotational motion is less noticeable compared to the translational motion especially for the off-diagonal component.
This observation can be explained by the fact that the rotational mobility corrections exhibit a faster decay with the distance from the membrane, scaling as $z_0^{-3}$ compared to $z_0^{-1}$ for translational motion.
Again, a good agreement is obtained for the rotational mobility corrections between analytical predictions and numerical simulations.

\section{Conclusions}\label{conclusions}

In this paper we have computed the leading-order translational, rotational and translation-rotation coupling hydrodynamic mobilities of an arbitrary shaped axisymmetric particle immersed in a Newtonian fluid in the vicinity of an elastic cell membrane.
The resulting equations contain (i) the particle-independent mulitpole elements of the near-membrane Green's integral operator which have been calculated in analytical form in the present work and (ii) the mobility tensor of the particle \textit{in bulk}.
The mobility corrections are frequency-dependent complex quantities due to the memory induced by the membrane.
They are expressed in terms of the particle orientation and two dimensionless parameters $\beta$ and $\betaB$ that account for the shearing and bending related contributions, respectively.
In the zero-frequency limit, or equivalently for infinite elastic and bending moduli, we recover the mobilities near a hard no-slip wall.
We apply our general formalism to a prolate spheroid and find very good agreement with numerical simulations performed for a truly extended spheroidal particle over the whole frequency spectrum.

\begin{acknowledgements}
ADMI and SG thank the Volkswagen Foundation for financial support and acknowledge the Gauss Center for Supercomputing e.V. for providing computing time on the GCS Supercomputer SuperMUC at Leibniz Supercomputing Center. 
This work has been supported by the Ministry of Science and Higher Education of Poland via the Mobility Plus Fellowship awarded to ML.
This article is based upon work from COST Action MP1305, supported by COST (European Cooperation in Science and Technology).
\end{acknowledgements}

\appendix

\section{The Green's function for an elastic membrane}\label{greens}

The Green's functions for an elastic membrane have been derived and discussed in detail in earlier papers \citep{daddi16,daddi16c}. 
Here, we only sketch the derivation which starts with a 2D Fourier transform of the Stokes equations and boundary conditions. 
It is convenient to introduce an orthogonal basis in the $xy$ plane, spanned by the unit vectors $\el=\vect{q}/|\vect{q}|$ and $\et=\ez\times\el$, respectively parallel and perpendicular to the wave vector~$\vect{q}$. 
This basis is rotated by the angle $\phi=\arctan (q_y/q_x)$ with respect to the laboratory frame.

{
After the pressure has been eliminated from the Fourier transformed momentum equations, the following set of ordinary differential equations is obtained
\begin{subequations}
      \begin{align}
       q^2 \vt_t - \vt_{t,zz} &= \frac{\Ft_t}{\eta} \delta (z-z_0) \, , \label{transverseEquation}\\
      \vt_{z,zzzz} - 2q^2 \vt_{z,zz} + q^4 \vt_z &= \frac{q^2 \Ft_z}{\eta} \delta(z-z_0) + \frac{iq \Ft_l}{\eta} \delta' (z-z_0) \, , \label{normalEquation}\\
       \vt_l &= \frac{i \vt_{z,z}}{q} \, , \label{longitudinalFromNormal}
      \end{align}
      \label{transformedEquations}
\end{subequations}
where $\delta'$ is the derivative of the Dirac delta function.
After some algebra, it can be shown that the traction jump due to shearing as stated in Eq. \eqref{tangentialCondition} imposes at $z=0$ the following discontinuities
\begin{equation}
   [\vt_{t,z}] = \left. -iB\alpha  q^2 \vt_t \right|_{z = 0} \, , \quad
   ~[\vt_{z,zz}] = \left. -{4i\alpha q^2} \vt_{z,z} \right|_{z = 0} \, ,
   \label{tangentialConditionFinal}
\end{equation}
where $\alpha := \kS/3 B\eta\omega$ is a characteristic length for shear and area dilatation with $B := 2/(1+C)$. 
The normal traction jump as given by Eq. \eqref{normalCondition} leads to
\begin{equation}
 [\vt_{z,zzz}] = \left. 4i \alphaB^3 q^6 \vt_z \right|_{z = 0} \, ,\label{normalConditionFinal}
\end{equation}
where $\alphaB^3 := \kB /4\eta\omega$, with $\alphaB$ being a characteristic length for bending.
The dimensionless numbers $\beta$ and $\betaB$ stated in Eq.~\eqref{beta_and_betaB_def} are defined as $\beta:=2z_0/\alpha$ and $\betaB:=2z_0/\alphaB$.}

The Green's tensor in this basis $\{\el,\et,\ez\}$ has the form 
\begin{equation}\label{GT}
 \boldsymbol{\Gt} (q,z,\omega) = \left(
		    \begin{array}{ccc}
		     \Gt_{ll} & 0 & \Gt_{lz} \\
		     0 & \Gt_{tt} & 0 \\
		     \Gt_{zl} & 0 & \Gt_{zz}
		    \end{array}
		    \right) \, .
\end{equation}

The components of the Green's functions for $z \ge 0$ are expressed by
\begin{align}
 \Gt_{zz}   &= \frac{1}{4 \eta q} 
 \bigg(
 \left( 1+q|z - z_0| \right) \ue^{-q|z-z_0|} 
   + \left( \frac{\ui\alpha z z_0 q^3}{1-\ui\alpha q} + \frac{\ui\alphaB^3 q^3 (1+qz)(1+q z_0)}{1-\ui\alphaB^3 q^3} \right) \ue^{-q(z+z_0)}  
 \bigg) \, ,
 \nonumber
 \\
  \Gt_{ll}  &= \frac{1}{4 \eta q} 
 \bigg(
(1-q |z - z_0|) \ue^{-q|z-z_0|} 
  + \left( \frac{\ui\alpha q (1-q z_0)(1-qz)}{1-\ui\alpha q} + \frac{\ui z z_0 \alphaB^3 q^5}{1-\ui \alphaB^3 q^3} \right) \ue^{-q(z+z_0)}  
 \bigg) \, , 
 \nonumber
 \\
 \Gt_{tt}  &= \frac{1}{2 \eta q} \bigg( \ue^{-q|z-z_0|}
  + \frac{\ui B\alpha q}{2-\ui B \alpha q} \ue^{-q(z+z_0)}  \bigg) \, ,
 \nonumber
\end{align}
with the off-diagonal components 
\begin{align}
 \Gt_{lz}   &= \frac{\ui}{4 \eta q} 
 \bigg(
-q (z - z_0) \ue^{-q|z-z_0|}
  + \left( \frac{\ui\alpha z_0 q^2 (1-qz)}{1-\ui\alpha q} - \frac{\ui\alphaB^3 z q^4 (1+q z_0)}{1-\ui \alphaB^3 q^3} \right) \ue^{-q(z+z_0)}  
 \bigg) \, , \nonumber \\
 \Gt_{zl}  &= \frac{{\ui}}{4 \eta q} 
 \bigg(
 -q(z - z_0) \ue^{-q|z-z_0|}
  + \left(- \frac{\ui\alpha z q^2 (1-q z_0)}{1-\ui\alpha q} + \frac{\ui \alphaB^3 q^4 z_0 (1+qz)}{1-\ui \alphaB^3 q^3} \right) \ue^{-q(z+z_0)}  
 \bigg) \, . \nonumber
\end{align}
The terms which contain $\ue^{-q|z-z_0|}$ are the Fourier-transformed elements of the Oseen tensor and do not depend on the elastic properties of the membrane. 
The remaining part comes from interactions with the interface. 
We now back-transform \eqref{GT} to the laboratory frame. Defining
\begin{equation}
 \Gt_{\pm} (q,z,\omega) := \Gt_{tt}(q,z,\omega) \pm \Gt_{ll}(q,z,\omega) \, , \nonumber
\end{equation}
and performing the inverse spatial Fourier transform \citep{bracewell99}, we find that the Green's functions for a point force acting at $\vect{r}_0 = (0,0,z_0)$ can be presented in terms of the following convergent infinite integrals 
\begin{align}
{\G}_{zz} (\R, \omega) &= \frac{1}{2\upi}
  \int_{0}^{\infty}  \Gt_{zz} (q,z,z_0,\omega) J_0 (\rho q) q \Intd q \, , \nonumber \\
\G_{xx} (\R, \omega) &= \frac{1}{4\upi}  \int_0^{\infty} \bigg( \Gt_{+} (q,z,z_0,\omega) J_0 (\rho q) 
                     +   \Gt_{-} (q,z,z_0,\omega) J_2 (\rho q) \cos 2\Theta \bigg) q \Intd q \, ,  \nonumber \\
\G_{yy} (\R,\omega)  &= \frac{1}{4\upi}  \int_0^{\infty} \bigg( \Gt_{+} (q,z,z_0,\omega) J_0 (\rho q) 
                     - \Gt_{-} (q,z,z_0,\omega) J_2 (\rho q) \cos 2\Theta \bigg) q \Intd q \, ,  \nonumber \\
\G_{xy} (\R, \omega) &= \frac{\sin 2\Theta}{4\upi} \int_0^\infty \Gt_{-} (q,z,z_0,\omega) J_2 (\rho q) q \Intd q \, , \nonumber \\
\G_{rz} (\R, \omega) &= \frac{\ui}{2\upi} \int_{0}^{\infty} \Gt_{lz} (q,z,z_0,\omega) J_1 (\rho q) q \Intd q \, , \nonumber \\
\G_{zr} (\R, \omega) &= \frac{\ui}{2\upi} \int_{0}^{\infty} \Gt_{zl} (q,z,z_0,\omega) J_1 (\rho q) q \Intd q \, , \nonumber
\end{align}
where $\rho:= \sqrt{x^2 + y^2}$ is the radial distance from the origin, and $\Theta := \arctan (y/x)$ is the angle formed by the radial and $x$ axis.
Furthermore, $\G_{xz}=\G_{rz}\cos\Theta$, $\G_{yz}=\G_{rz}\sin\Theta$, $\G_{zx}=\G_{zr}\cos\Theta$, $\G_{zy}=\G_{zr}\sin\Theta$ and $\G_{yx} = \G_{xy}$.
Here $J_n$ denotes the Bessel function \citep{abramowitz72} of the first kind of order~$n$.

In the vanishing frequency limit, or equivalently for infinite membrane shearing and bending rigidities, the well-known Blake tensor \citep{blake71} is recovered for all the components of the Green's functions.

 
\section{Derivation of general mobility relations}
\label{appInversion}

Here we sketch the manipulations that lead from the corrected friction tensor, given by Eqs. \eqref{corrtt} through \eqref{corrrr}, to the mobility correction in Eqs. \eqref{Mtt} through \eqref{Mrr}. 
We shall focus on the $tt$ part only, since the others follow analogously.  Relation \eqref{frictionmobility}, rewritten as $\mi\bzz = \bm{1}$, defines the relations between elements of the friction and mobility tensors of a particle close to a membrane. 
The membrane-corrected $tt$ friction tensor and the membrane-corrected $tt$ mobility are thus related by
\begin{align} 
\mi^{tt}\bzz^{tt}+\mi^{tr}\bzz^{rt} &= \bm{1}, \\
\mi^{tt}\bzz^{tr}+\mi^{tr}\bzz^{rr} &= \bm{0}
\end{align}
from which we have
\begin{equation}\label{mittinv}
\mi^{tt} = [\bzz^{tt} - \bzz^{tr}(\bzz^{rr})^{-1}\bzz^{rt}]^{-1}.
\end{equation}
We know from Eqs. \eqref{corrtt}-\eqref{corrrr} that the corrected friction has the following structure
\begin{align} 
\bzz^{tt} &= \bzz^{tt}_0 + \bs{\Delta}\bzz^{tt}, \\  
\bzz^{rr} &= \bzz^{rr}_0 + \bs{\Delta}\bzz^{rr}, \\
\bzz^{tr} &= \bs{\Delta}\bzz^{tr}, \\
\bzz^{rt} &= \bs{\Delta}\bzz^{rt}
\end{align}
with the known distance-dependence of these elements. Moreover, for an axially symmetric particle, we have
\begin{equation} 
\bzz^{tt}_0 = (\mi^{tt}_0)^{-1}, \qquad\qquad \bzz^{rr}_0 = (\mi^{rr}_0)^{-1},
\end{equation}
since the bulk friction and mobility tensors are diagonal. We now rewrite Eq. \eqref{mittinv} as
\begin{equation} 
\mi^{tt} = \mi_0^{tt} [\bs{1} + (\bs{\Delta}\bzz^{tt})\mi^{tt}_0 - (\bs{\Delta}\bzz^{tr})(\bzz_0^{rr}+\bs{\Delta}\bzz^{rr})^{-1}(\bs{\Delta}\bzz^{rt})\mi^{tt}_0]^{-1}
\end{equation}
 and expand the expression $1/(1+\delta) = 1-\delta+\delta^2 -\ldots$ around the bulk quantities. 
 Restricting to quantities decaying slower than $z_0^{-3}$, we immediately find Eq. \eqref{Mtt}. 
An analogous procedure leads to the $tr$, $rt$ and $rr$ mobilities, where the elements of the bulk friction and mobility tensors combine to contribute only in the form of $\mi_0^{dr}=\mi^{rr}_0 \bzz^{rd}_0$ and $\mi^{rd}_0 = -\bzz^{dr}_0 \mi^{rr}_0$. 
The latter relations follow from the definitions \eqref{friction_single} and \eqref{mobility_single}. 

 
\section{Expressions required for the spheroid mobilities}\label{analyticalExpressions}

The results for the correction are given in terms of the wall-particle distance $z_0$, its inclination angle $\theta$ and functions denoted by capital letters in Eqs.~\eqref{deltaMu_tt_11} through \eqref{deltaMu_rr_33} of the dimensionless shearing and bending parameters, $\beta$ and $\betaB$. 
Below, we provide explicit expressions for these functions. 
They can conveniently be expressed in terms of higher order exponential integrals \citep{abramowitz72}.
The contributions from the membrane shearing (index $\mathrm{S}$) and bending (index $\mathrm{B}$) are given separately. 
By summing up both, we arrive at the final expressions. Notably, in the limit of vanishing frequency, our results are in complete agreement with those given by \citep{lisicki15thesis,lisicki16}.

\subsection{Translational mobility}

For the functions $P$ and $Q$, we find the shearing contribution as
 \begin{align} \nonumber
  P_\mathrm{S} (\beta) &= -\frac{5}{4}+\frac{\beta^2}{8}-\frac{3\ui\beta}{8} + \frac{2\ui\beta}{B} \Gamma_2 
                       + \left( -\frac{\beta^2}{2}+\frac{\ui\beta}{2} \left( 1-\frac{\beta^2}{4} \right) \right) \ue^{\ui\beta} \E_1 (\ui\beta) \, ,  \nonumber \\
  Q_\mathrm{S} (\beta) &= -\frac{3}{2}  \ue^{\ui\beta} \E_4(\ui\beta) \, ,  \nonumber
 \end{align}
and the bending part 
 \begin{align} \nonumber
  P_\mathrm{B} (\betaB) &= -\frac{1}{4} + \frac{\ui\betaB^3}{24} \left( \phi_{+} + \Gamma_\mathrm{B} \right) \, ,  \\ \nonumber
  Q_\mathrm{B} (\betaB) &= -\frac{5}{2} + \ui\betaB \bigg( 
			  \left( \frac{\betaB^2}{12}+\frac{\ui\betaB}{6}+\frac{1}{6} \right) \phi_{+} 
			  + \left( \frac{\betaB^2}{12} - \frac{\ui\betaB}{3} - \frac{1}{3} \right) \ue^{-\ui\betaB} \E_1(-\ui\betaB) \nonumber \\
			  &+\frac{\sqrt{3}}{6}(\betaB+\ui)\phi_{-}
			  \bigg) \, ,  \nonumber
 \end{align}
with
 \begin{align}
  \phi_{\pm} &= \ue^{-\ui\overline{\zB}} \E_1(-\ui\overline{\zB}) \pm \ue^{-\ui\zB} \E_1(-\ui\zB) \, ,  \nonumber \\
  \Gamma_2 &= \ue^{\frac{2\ui\beta}{B}} \E_1 \left( \frac{2\ui\beta}{B} \right) \, ,  \nonumber \\
  \Gamma_\mathrm{B} &= \ue^{-\ui\betaB} \E_1(-\ui\betaB) \, , {\nonumber} 
 \end{align}
where $\zB := \betaB \ue^{2\ui\upi/3}$ and the bar denotes the complex conjugate. The function $\mathrm{E}_n$ is the generalised exponential integral, $\mathrm{E}_n(x)=\int_{1}^{\infty} t^{-n} \ue^{-xt}\mathrm{d}t$.

\subsection{Translation-rotation coupling}

The translation-rotation coupling is determined by the functions $M$ and $N$, which we similarly decompose into two parts. Recalling that $B=2/(1+C)$, the shearing part reads
 \begin{align}
  M_{\mathrm{S}} (\beta) &= \frac{3}{4}-\ui\beta \left( \frac{1}{4}+\frac{1}{B} \right) +\frac{3\beta^2}{8}+\frac{\ui\beta^3}{8} 
                         - \frac{\beta^2}{2} \left( 1+\ui\beta-\frac{\beta^2}{4} \right)  \Gamma_1 -\frac{2\beta^2}{B^2} \Gamma_2 \, , {\nonumber} \\
  N_{\mathrm{S}} (\beta) &= -\frac{3}{4} +\ui\beta \left( \frac{1}{2}+\frac{2}{B} \right) -\frac{3\beta^2}{8} + \frac{\ui\beta^3}{8} 
                         + {\beta^2}\left( 1+\frac{\ui\beta}{4} + \frac{\beta^2}{8} \right)  \Gamma_1 + \frac{4\beta^2}{B^2} \Gamma_2 \, ,  {\nonumber}
 \end{align}
while the bending part is
 \begin{align}
  M_{\mathrm{B}} (\betaB) &= \frac{3}{4} -\frac{\ui\betaB^3}{8} + \frac{\betaB^4}{24} \Gamma_\mathrm{B} + \frac{\betaB^3}{24} \psi \, , {\nonumber} \\
  N_{\mathrm{B}} (\betaB) &= \frac{9}{4} -\frac{\ui\betaB^3}{8} - \frac{\betaB^3}{4} \left(\ui-\frac{\betaB}{6} \right) \Gamma_\mathrm{B} + \frac{\betaB^3}{24} \psi - \frac{\ui\betaB^3}{4} \phi_{+} \, , {\nonumber}
 \end{align}
where we defined
 \begin{align}
  \Gamma_1 &= \ue^{\ui\beta} \E_1(\ui\beta) \, , \nonumber \\
  \psi &= \zBbar \ue^{-\ui\zBbar} \E_1 (-\ui\zBbar) + \zB \ue^{-\ui\zB} \E_1 (-\ui\zB) \, . {\nonumber}
 \end{align}


\subsection{Rotational mobility}

The rotational mobility is described by a set of functions. 
The functions $A_0$ and $A_2$ defined for the component $\Delta \mu_{11}^{rr}$ in Eq.~\eqref{deltaMu_rr_11} are given by 

 \begin{align}
  A_{0, \mathrm{S}} (\beta) &= -\frac{3}{2}+\frac{\ui\beta}{2}\left(1+\frac{1}{B} \right)+\beta^2 \left( \frac{1}{2}+\frac{1}{B^2} \right) -\frac{\ui\beta^3}{2} \Gamma_1 - \frac{2\ui\beta^3}{B^3} \Gamma_2 \, , {\nonumber} \\ 
  A_{2, \mathrm{S}} (\beta) &= \frac{1}{2} +\frac{\ui\beta}{2} \left( \frac{1}{B}-1 \right)+\beta^2\left( \frac{1}{B^2}-\frac{1}{2} \right) +\frac{i\beta^3}{2} \Gamma_1 - \frac{2\ui\beta^3}{B^3} \Gamma_2 \, , {\nonumber} \\
  A_{0, \mathrm{B}} (\betaB) &= -A_{2, \mathrm{B}} (\betaB) = -1+\frac{\ui\betaB^3}{6} \left( \phi_{+} + \Gamma_\mathrm{B} \right) \, . {\nonumber}
 \end{align}


For the component $\Delta \mu_{13}^{rr}$, the function $D$ defined in Eq.~\eqref{deltaMu_rr_13} is given by
 \begin{align}
  D_{\mathrm{S}} (\beta) &= -\frac{1}{2}+\frac{\ui\beta}{2} \left( 1-\frac{1-\lambda}{B} \right) 
                          + \beta^2 \left( \frac{1}{2}+\frac{\lambda}{4} - \frac{1-\lambda}{B^2} \right) 
			  +\ui\beta^3 \bigg( \frac{2(1-\lambda)}{B^3} \Gamma_2 \nonumber \\
			  &- \frac{\lambda+1}{2}\Gamma_1 
			  + \frac{\lambda}{4} \bigg) 
  + \frac{\lambda \beta^4}{4} \Gamma_1 \, , {\nonumber} \\
  D_{\mathrm{B}} (\betaB) &= -1+\frac{3\lambda}{2} + \frac{\betaB^3 }{4} \left( -\ui\lambda+\frac{2i}{3}\Gamma_\mathrm{B} + \frac{2i\phi_{+}+\lambda\psi}{3} \right) 
                          +\frac{\lambda \betaB^4}{12} \Gamma_\mathrm{B} \, . {\nonumber}
 \end{align}


Further, the shearing related parts of $C_0$, $C_2$ and $C_4$ as defined for the correction $\Delta\mu_{22}^{rr}$ in Eq.~\eqref{deltaMu_rr_22} read
 \begin{align}
  C_{0,\mathrm{S}}  (\beta) &= -\frac{3}{2}\left(1+\lambda^2\right) + \left( \frac{\ui\lambda^2\beta^2}{8}-\frac{\lambda(1-\lambda)\beta}{2}-\frac{\ui(1-\lambda)^2}{2} \right) \beta^3 \Gamma_1-\frac{2\ui}{B^3}(1-\lambda)^2 \beta^3 \Gamma_2 \nonumber \\
  &+\frac{\ui\beta}{4} \left( \lambda^2+2+\frac{2(1-\lambda)^2}{B} \right)
  +\frac{\beta^2}{4} \left( \lambda^2-2\lambda+2+\frac{4(1-\lambda)^2}{B^2} \right) \nonumber \\
  &- \ui\lambda \beta^3 \left( \frac{1}{2} - \frac{3\lambda}{8} \right) - \frac{\lambda^2 \beta^4}{8} \, , {\nonumber} \\
  C_{2,\mathrm{S}}  (\beta) &= \frac{3\lambda}{4} - \beta^3(2+\ui\beta) \left( \ui-\frac{3}{32}\lambda(\beta-2i) \right) \Gamma_1 - \frac{2i}{B^3}(4-3\lambda) \beta^3 \Gamma_2
  +\frac{\ui\beta}{2} \left( \frac{3\lambda}{8} + \frac{4-3\lambda}{B} \right) \nonumber \\
  &+\beta^2 \left( \frac{3\lambda}{16}+1+\frac{4-3\lambda}{B^2} \right)
  +\ui\beta^3 \left( \frac{9\lambda}{32} + 1 \right) - \frac{3\lambda}{32} \beta^4 \, , {\nonumber} \\
  C_{4, \mathrm{S}}  (\beta) &= -\frac{3}{4} + 3\ui \beta \left( -\frac{1}{16}+\frac{1}{2B} \right) 
  +3\beta^2  \left(-\frac{1}{16} + \frac{1}{B^2}  \right)
  +\ui\beta^3  \left( -\frac{9}{32}-\frac{6}{B^3} \Gamma_2 \right) \nonumber \\
  &+\frac{3}{8} \beta^3\left( \ui-\beta-\frac{\ui}{4} \beta^2  \right) \Gamma_1
  +\frac{3}{32}  \beta^4 \, ,  {\nonumber}
  \end{align}
and the bending-related parts read
 \begin{align}
  C_{0, \mathrm{B}}  (\betaB) &= -1-3\lambda(1+\lambda)+ \frac{\ui\betaB^3}{8} \lambda (\lambda+4)
  -\frac{\betaB^3}{24} (\ui\lambda^2\betaB^2+4\lambda\betaB-4\ui) \Gamma_\mathrm{B} \nonumber \\
   &+\frac{\betaB^3}{6} \left( \ui\phi_{+}-\lambda\psi + \frac{\ui\lambda^2}{4} \betaB (\psi+\betaB\phi_{+}) \right), {\nonumber} \\
  C_{2, \mathrm{B}}  (\betaB) &= 6-\frac{9}{4}\lambda-\ui\betaB^3\left( 1-\frac{3\lambda}{32} \right)
  +\left( \frac{1}{3} - \frac{\ui\lambda\betaB}{32} \right) \betaB^4 \Gamma_\mathrm{B}
  +\frac{\betaB^3}{3} \psi + \frac{\ui\lambda \betaB^4}{32} (\psi+\betaB \phi_{+}) \, , {\nonumber} \\
  C_{4, \mathrm{B}}  (\betaB) &= \frac{9}{4}-\frac{3\ui\betaB^3}{32}+\frac{\ui\betaB^5}{32} \Gamma_\mathrm{B}-\frac{\ui\betaB^4}{32}(\psi+\betaB\phi_{+}) \, , {\nonumber}
 \end{align}
 

Finally, the functions $H_0$ and $H_2$ defined for the component $\Delta\mu_{33}^{rr}$ in Eq.~\eqref{deltaMu_rr_33} read
 \begin{align}
  H_{0, \mathrm{S}}  (\beta) &= -1-\frac{3}{4} \lambda^2 + \ui\beta \left( \frac{\lambda^2+2}{2B} + \frac{\lambda^2}{16} \right)
  +\beta^2 \left( \frac{2+\lambda^2}{B^2} + \frac{\lambda^2}{16} \right) 
  +\frac{\lambda^2\beta^3}{32} (\ui\beta^2-4\ui+4\beta) \Gamma_1
  \nonumber \\
  &+ \beta^3 \left( \frac{3\ui\lambda^2}{32} - \frac{2\ui}{B^3}(\lambda^2+2) \Gamma_2 \right) - \frac{\lambda^2 \beta^4}{32} \, , {\nonumber} \\
  H_{2, \mathrm{S}}  (\beta) &= -\frac{1}{2}-\frac{3}{4}\lambda^2 + \beta \left( \frac{3\ui}{16}\lambda^2 +\frac{\ui\lambda}{B} + \frac{\ui}{2}-\frac{\ui}{2B} \right) 
  + \beta^2 \left( \frac{3}{16}\lambda^2 +\frac{\lambda+1}{2} + \frac{2\lambda}{B^2}  - \frac{1}{B^2} \right)  \nonumber \\
  &+\frac{\beta^3}{32} \left( {9\ui\lambda^2} +16{\ui\lambda} +\frac{64\ui(1-2\lambda)\Gamma_2}{B^3} + (4+6\lambda+3\ui\lambda\beta)(\lambda\beta-2\ui\lambda-4\ui)\Gamma_1 \right) -\frac{3\lambda^2 \beta^4}{32}  \, ,  {\nonumber}
 \end{align}
and 
 \begin{align}
  H_{0, \mathrm{B}}  (\betaB) &= \lambda^2 \left( -\frac{3}{4} + \frac{\ui\betaB^3}{32}-\frac{\ui\betaB^5}{96} \Gamma_\mathrm{B} + \frac{\ui \betaB^4}{96} (\psi + \betaB\phi_{+}) \right) \, , {\nonumber} \\
  H_{2, \mathrm{B}}  (\betaB) &= -1+3\lambda-\frac{9}{4}\lambda^2 + \betaB^3 \left(  \ui\lambda\left(\frac{3}{32}\lambda-\frac{1}{2}\right) 
  +\frac{1}{6} \left( \ui+\lambda\betaB-\frac{3\ui\lambda^2\betaB^2}{16} \right) \Gamma_\mathrm{B} \right) \nonumber \\
  &+\frac{\ui\lambda^2\betaB^4}{32} (\psi+\betaB\phi_{+})+\frac{\betaB^3}{6} (\lambda\psi+\ui\phi_{+}) \, . {\nonumber}
 \end{align}

\bibliographystyle{jfm}

\input{main.bbl}
\end{document}

%% file: commands.tex
\newcommand{\vect}[1]{\boldsymbol{#1}}

\newcommand{\boldNabla}{\boldsymbol{\nabla}}

\newcommand{\kS}{\kappa_\mathrm{S}}
\newcommand{\kA}{\kappa_\mathrm{A}}
\newcommand{\kB}{\kappa_\mathrm{B}}

\newcommand{\alphaB}{\alpha_\mathrm{B}}

\newcommand{\betaB}{\beta_\mathrm{B}}

\newcommand{\G}{\mathcal{G}}

\newcommand{\R}{\vect{r}}

\newcommand{\Intd}{\mathrm{d }}

\newcommand{\DeltaPara}{\Delta_{\parallel}}

\newcommand{\vt}{\tilde{v}}

\newcommand{\Gt}{\tilde{\G}}
\newcommand{\Ft}{\tilde{F}}

\newcommand{\E}{\operatorname{E}}

\newcommand{\Gmatr}{\boldsymbol{\mathcal{G}}}

\newcommand{\zB}{z_{\mathrm{B}}}
\newcommand{\zBbar}{\overline{z_{\mathrm{B}}}}

\newcommand{\ex}{{\boldsymbol{e}}_x}
\newcommand{\ey}{{\boldsymbol{e}}_y}
\newcommand{\ez}{{\boldsymbol{e}}_z}

\newcommand{\et}{{\boldsymbol{e}}_t}
\newcommand{\el}{{\boldsymbol{e}}_l}

\newcommand{\bs}[1]{\boldsymbol{#1}}

\newcommand{\br}{\boldsymbol{r}}
\newcommand{\bv}{\boldsymbol{v}}
\newcommand{\bg}{\boldsymbol{g}}

\newcommand{\BR}{\vect{r}_0}
\newcommand{\BF}{\boldsymbol{F}}
\newcommand{\BT}{\boldsymbol{T}}
\newcommand{\BS}{\boldsymbol{S}}
\newcommand{\BV}{\boldsymbol{V}}
\newcommand{\BE}{\boldsymbol{E}}

\newcommand{\bzz}{\bm{\zeta}}
\newcommand{\mi}{\bm{\mu}}

\newcommand{\oOmega}{\bm{\Omega}}

\newcommand{\uu}{{\boldsymbol{u}}}

\newcommand{\integr}[3]{\int_{#1}^{#2} \!\! \mathrm{d} {#3}\,}

\newcommand{\ui}{\mathrm{i}}
\newcommand{\ue}{\mathrm{e}}



%% file: Pics/spheroidNearMembrane.tex
\begingroup
  \makeatletter
  \providecommand\color[2][]{%
    \GenericError{(gnuplot) \space\space\space\@spaces}{%
      Package color not loaded in conjunction with
      terminal option `colourtext'%
    }{See the gnuplot documentation for explanation.%
    }{Either use 'blacktext' in gnuplot or load the package
      color.sty in LaTeX.}%
    \renewcommand\color[2][]{}%
  }%
  \providecommand\includegraphics[2][]{%
    \GenericError{(gnuplot) \space\space\space\@spaces}{%
      Package graphicx or graphics not loaded%
    }{See the gnuplot documentation for explanation.%
    }{The gnuplot epslatex terminal needs graphicx.sty or graphics.sty.}%
    \renewcommand\includegraphics[2][]{}%
  }%
  \providecommand\rotatebox[2]{#2}%
  \@ifundefined{ifGPcolor}{%
    \newif\ifGPcolor
    \GPcolorfalse
  }{}%
  \@ifundefined{ifGPblacktext}{%
    \newif\ifGPblacktext
    \GPblacktexttrue
  }{}%
  \let\gplgaddtomacro\g@addto@macro
  \gdef\gplbacktext{}%
  \gdef\gplfronttext{}%
  \makeatother
  \ifGPblacktext
    \def\colorrgb#1{}%
    \def\colorgray#1{}%
  \else
    \ifGPcolor
      \def\colorrgb#1{\color[rgb]{#1}}%
      \def\colorgray#1{\color[gray]{#1}}%
      \expandafter\def\csname LTw\endcsname{\color{white}}%
      \expandafter\def\csname LTb\endcsname{\color{black}}%
      \expandafter\def\csname LTa\endcsname{\color{black}}%
      \expandafter\def\csname LT0\endcsname{\color[rgb]{1,0,0}}%
      \expandafter\def\csname LT1\endcsname{\color[rgb]{0,1,0}}%
      \expandafter\def\csname LT2\endcsname{\color[rgb]{0,0,1}}%
      \expandafter\def\csname LT3\endcsname{\color[rgb]{1,0,1}}%
      \expandafter\def\csname LT4\endcsname{\color[rgb]{0,1,1}}%
      \expandafter\def\csname LT5\endcsname{\color[rgb]{1,1,0}}%
      \expandafter\def\csname LT6\endcsname{\color[rgb]{0,0,0}}%
      \expandafter\def\csname LT7\endcsname{\color[rgb]{1,0.3,0}}%
      \expandafter\def\csname LT8\endcsname{\color[rgb]{0.5,0.5,0.5}}%
    \else
      \def\colorrgb#1{\color{black}}%
      \def\colorgray#1{\color[gray]{#1}}%
      \expandafter\def\csname LTw\endcsname{\color{white}}%
      \expandafter\def\csname LTb\endcsname{\color{black}}%
      \expandafter\def\csname LTa\endcsname{\color{black}}%
      \expandafter\def\csname LT0\endcsname{\color{black}}%
      \expandafter\def\csname LT1\endcsname{\color{black}}%
      \expandafter\def\csname LT2\endcsname{\color{black}}%
      \expandafter\def\csname LT3\endcsname{\color{black}}%
      \expandafter\def\csname LT4\endcsname{\color{black}}%
      \expandafter\def\csname LT5\endcsname{\color{black}}%
      \expandafter\def\csname LT6\endcsname{\color{black}}%
      \expandafter\def\csname LT7\endcsname{\color{black}}%
      \expandafter\def\csname LT8\endcsname{\color{black}}%
    \fi
  \fi
  \setlength{\unitlength}{0.0500bp}%
  \begin{picture}(3600.00,2520.00)%
    \gplgaddtomacro\gplbacktext{%
      \csname LTb\endcsname%
      \put(2126,919){\makebox(0,0)[l]{\strut{}$z_0$}}%
      \put(1228,1278){\makebox(0,0)[l]{\strut{}$a$}}%
      \put(1587,1529){\makebox(0,0)[l]{\strut{}$c$}}%
      \put(2844,703){\makebox(0,0)[l]{\strut{}$x$}}%
      \put(1946,2176){\makebox(0,0)[l]{\strut{}$\theta$}}%
      \put(2126,1601){\makebox(0,0)[l]{\strut{}$\vect{u}_1$}}%
      \put(869,1637){\makebox(0,0)[l]{\strut{}$\vect{u}_3$}}%
    }%
    \gplgaddtomacro\gplfronttext{%
    }%
    \gplbacktext
    \put(0,0){\includegraphics{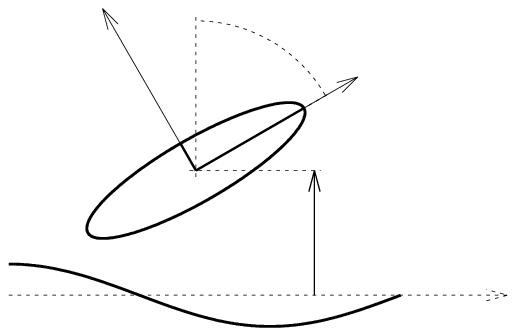}}%
    \gplfronttext
  \end{picture}%
\endgroup

%% file: Pics/DeltaMu_Tran.tex
\begingroup
  \makeatletter
  \providecommand\color[2][]{%
    \GenericError{(gnuplot) \space\space\space\@spaces}{%
      Package color not loaded in conjunction with
      terminal option `colourtext'%
    }{See the gnuplot documentation for explanation.%
    }{Either use 'blacktext' in gnuplot or load the package
      color.sty in LaTeX.}%
    \renewcommand\color[2][]{}%
  }%
  \providecommand\includegraphics[2][]{%
    \GenericError{(gnuplot) \space\space\space\@spaces}{%
      Package graphicx or graphics not loaded%
    }{See the gnuplot documentation for explanation.%
    }{The gnuplot epslatex terminal needs graphicx.sty or graphics.sty.}%
    \renewcommand\includegraphics[2][]{}%
  }%
  \providecommand\rotatebox[2]{#2}%
  \@ifundefined{ifGPcolor}{%
    \newif\ifGPcolor
    \GPcolorfalse
  }{}%
  \@ifundefined{ifGPblacktext}{%
    \newif\ifGPblacktext
    \GPblacktexttrue
  }{}%
  \let\gplgaddtomacro\g@addto@macro
  \gdef\gplbacktext{}%
  \gdef\gplfronttext{}%
  \makeatother
  \ifGPblacktext
    \def\colorrgb#1{}%
    \def\colorgray#1{}%
  \else
    \ifGPcolor
      \def\colorrgb#1{\color[rgb]{#1}}%
      \def\colorgray#1{\color[gray]{#1}}%
      \expandafter\def\csname LTw\endcsname{\color{white}}%
      \expandafter\def\csname LTb\endcsname{\color{black}}%
      \expandafter\def\csname LTa\endcsname{\color{black}}%
      \expandafter\def\csname LT0\endcsname{\color[rgb]{1,0,0}}%
      \expandafter\def\csname LT1\endcsname{\color[rgb]{0,1,0}}%
      \expandafter\def\csname LT2\endcsname{\color[rgb]{0,0,1}}%
      \expandafter\def\csname LT3\endcsname{\color[rgb]{1,0,1}}%
      \expandafter\def\csname LT4\endcsname{\color[rgb]{0,1,1}}%
      \expandafter\def\csname LT5\endcsname{\color[rgb]{1,1,0}}%
      \expandafter\def\csname LT6\endcsname{\color[rgb]{0,0,0}}%
      \expandafter\def\csname LT7\endcsname{\color[rgb]{1,0.3,0}}%
      \expandafter\def\csname LT8\endcsname{\color[rgb]{0.5,0.5,0.5}}%
    \else
      \def\colorrgb#1{\color{black}}%
      \def\colorgray#1{\color[gray]{#1}}%
      \expandafter\def\csname LTw\endcsname{\color{white}}%
      \expandafter\def\csname LTb\endcsname{\color{black}}%
      \expandafter\def\csname LTa\endcsname{\color{black}}%
      \expandafter\def\csname LT0\endcsname{\color{black}}%
      \expandafter\def\csname LT1\endcsname{\color{black}}%
      \expandafter\def\csname LT2\endcsname{\color{black}}%
      \expandafter\def\csname LT3\endcsname{\color{black}}%
      \expandafter\def\csname LT4\endcsname{\color{black}}%
      \expandafter\def\csname LT5\endcsname{\color{black}}%
      \expandafter\def\csname LT6\endcsname{\color{black}}%
      \expandafter\def\csname LT7\endcsname{\color{black}}%
      \expandafter\def\csname LT8\endcsname{\color{black}}%
    \fi
  \fi
  \setlength{\unitlength}{0.0500bp}%
  \begin{picture}(11520.00,9072.00)%
    \gplgaddtomacro\gplbacktext{%
      \csname LTb\endcsname%
      \put(946,5240){\makebox(0,0)[r]{\strut{}-0.4}}%
      \put(946,5953){\makebox(0,0)[r]{\strut{}-0.3}}%
      \put(946,6666){\makebox(0,0)[r]{\strut{}-0.2}}%
      \put(946,7380){\makebox(0,0)[r]{\strut{}-0.1}}%
      \put(946,8093){\makebox(0,0)[r]{\strut{} 0}}%
      \put(946,8806){\makebox(0,0)[r]{\strut{} 0.1}}%
      \put(1078,5020){\makebox(0,0){\strut{}$10^{-6}$}}%
      \put(2149,5020){\makebox(0,0){\strut{}$10^{-4}$}}%
      \put(3221,5020){\makebox(0,0){\strut{}$10^{-2}$}}%
      \put(4292,5020){\makebox(0,0){\strut{}$10^{0}$}}%
      \put(5363,5020){\makebox(0,0){\strut{}$10^{2}$}}%
      \put(176,7023){\rotatebox{-270}{\makebox(0,0){\strut{}$\left. \Delta\mu_{33}^{tt} \middle/ \mu_\perp^t \right.$}}}%
      \put(3220,4690){\makebox(0,0){\strut{}$\beta$}}%
      \put(262,8806){\makebox(0,0){\large $a)$}}%
    }%
    \gplgaddtomacro\gplfronttext{%
    }%
    \gplgaddtomacro\gplbacktext{%
      \csname LTb\endcsname%
      \put(6838,5240){\makebox(0,0)[r]{\strut{}-0.04}}%
      \put(6838,6132){\makebox(0,0)[r]{\strut{} 0}}%
      \put(6838,7023){\makebox(0,0)[r]{\strut{} 0.04}}%
      \put(6838,7915){\makebox(0,0)[r]{\strut{} 0.08}}%
      \put(6838,8806){\makebox(0,0)[r]{\strut{} 0.12}}%
      \put(6970,5020){\makebox(0,0){\strut{}$10^{-6}$}}%
      \put(8008,5020){\makebox(0,0){\strut{}$10^{-4}$}}%
      \put(9046,5020){\makebox(0,0){\strut{}$10^{-2}$}}%
      \put(10084,5020){\makebox(0,0){\strut{}$10^{0}$}}%
      \put(11122,5020){\makebox(0,0){\strut{}$10^{2}$}}%
      \put(5936,7023){\rotatebox{-270}{\makebox(0,0){\strut{}$\left. \Delta\mu_{13}^{tt} \middle/ \sqrt{\mu_{\parallel}^t \mu_{\perp}^t} \right. = \left. \Delta\mu_{31}^{tt} \middle/ \sqrt{\mu_{\parallel}^t \mu_{\perp}^t} \right.$~~~~}}}%
      \put(9046,4690){\makebox(0,0){\strut{}$\beta$}}%
      \put(6088,8806){\makebox(0,0){\large $b)$}}%
    }%
    \gplgaddtomacro\gplfronttext{%
    }%
    \gplgaddtomacro\gplbacktext{%
      \csname LTb\endcsname%
      \put(946,704){\makebox(0,0)[r]{\strut{}-0.3}}%
      \put(946,1596){\makebox(0,0)[r]{\strut{}-0.2}}%
      \put(946,2487){\makebox(0,0)[r]{\strut{}-0.1}}%
      \put(946,3379){\makebox(0,0)[r]{\strut{} 0}}%
      \put(946,4271){\makebox(0,0)[r]{\strut{} 0.1}}%
      \put(1078,484){\makebox(0,0){\strut{}$10^{-6}$}}%
      \put(2149,484){\makebox(0,0){\strut{}$10^{-4}$}}%
      \put(3221,484){\makebox(0,0){\strut{}$10^{-2}$}}%
      \put(4292,484){\makebox(0,0){\strut{}$10^{0}$}}%
      \put(5363,484){\makebox(0,0){\strut{}$10^{2}$}}%
      \put(176,2487){\rotatebox{-270}{\makebox(0,0){\strut{}$\left. \Delta\mu_{11}^{tt} \middle/ \mu_\parallel^t \right.$}}}%
      \put(3220,154){\makebox(0,0){\strut{}$\beta$}}%
      \put(262,4271){\makebox(0,0){\large $c)$}}%
    }%
    \gplgaddtomacro\gplfronttext{%
    }%
    \gplgaddtomacro\gplbacktext{%
      \csname LTb\endcsname%
      \put(6706,704){\makebox(0,0)[r]{\strut{}-0.2}}%
      \put(6706,1893){\makebox(0,0)[r]{\strut{}-0.1}}%
      \put(6706,3082){\makebox(0,0)[r]{\strut{} 0}}%
      \put(6706,4271){\makebox(0,0)[r]{\strut{} 0.1}}%
      \put(6838,484){\makebox(0,0){\strut{}$10^{-6}$}}%
      \put(7909,484){\makebox(0,0){\strut{}$10^{-4}$}}%
      \put(8980,484){\makebox(0,0){\strut{}$10^{-2}$}}%
      \put(10051,484){\makebox(0,0){\strut{}$10^{0}$}}%
      \put(11122,484){\makebox(0,0){\strut{}$10^{2}$}}%
      \put(5936,2487){\rotatebox{-270}{\makebox(0,0){\strut{}$\left. \Delta\mu_{22}^{tt} \middle/ \mu_\perp^t \right.$}}}%
      \put(8980,154){\makebox(0,0){\strut{}$\beta$}}%
      \put(6022,4271){\makebox(0,0){\large $d)$}}%
    }%
    \gplgaddtomacro\gplfronttext{%
    }%
    \gplbacktext
    \put(0,0){\includegraphics{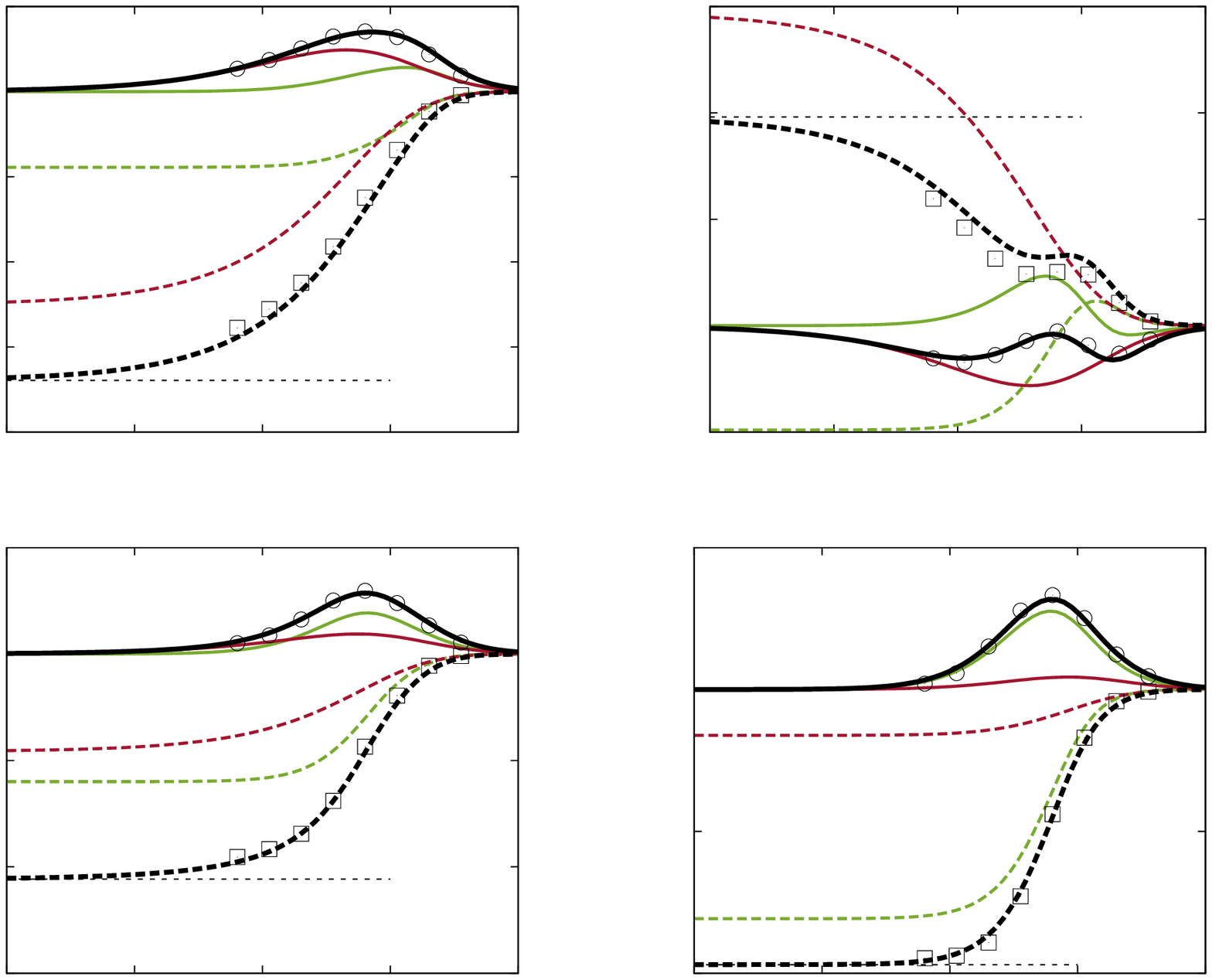}}%
    \gplfronttext
  \end{picture}%
\endgroup

%% file: Pics/DeltaMu_Coup.tex
\begingroup
  \makeatletter
  \providecommand\color[2][]{%
    \GenericError{(gnuplot) \space\space\space\@spaces}{%
      Package color not loaded in conjunction with
      terminal option `colourtext'%
    }{See the gnuplot documentation for explanation.%
    }{Either use 'blacktext' in gnuplot or load the package
      color.sty in LaTeX.}%
    \renewcommand\color[2][]{}%
  }%
  \providecommand\includegraphics[2][]{%
    \GenericError{(gnuplot) \space\space\space\@spaces}{%
      Package graphicx or graphics not loaded%
    }{See the gnuplot documentation for explanation.%
    }{The gnuplot epslatex terminal needs graphicx.sty or graphics.sty.}%
    \renewcommand\includegraphics[2][]{}%
  }%
  \providecommand\rotatebox[2]{#2}%
  \@ifundefined{ifGPcolor}{%
    \newif\ifGPcolor
    \GPcolorfalse
  }{}%
  \@ifundefined{ifGPblacktext}{%
    \newif\ifGPblacktext
    \GPblacktexttrue
  }{}%
  \let\gplgaddtomacro\g@addto@macro
  \gdef\gplbacktext{}%
  \gdef\gplfronttext{}%
  \makeatother
  \ifGPblacktext
    \def\colorrgb#1{}%
    \def\colorgray#1{}%
  \else
    \ifGPcolor
      \def\colorrgb#1{\color[rgb]{#1}}%
      \def\colorgray#1{\color[gray]{#1}}%
      \expandafter\def\csname LTw\endcsname{\color{white}}%
      \expandafter\def\csname LTb\endcsname{\color{black}}%
      \expandafter\def\csname LTa\endcsname{\color{black}}%
      \expandafter\def\csname LT0\endcsname{\color[rgb]{1,0,0}}%
      \expandafter\def\csname LT1\endcsname{\color[rgb]{0,1,0}}%
      \expandafter\def\csname LT2\endcsname{\color[rgb]{0,0,1}}%
      \expandafter\def\csname LT3\endcsname{\color[rgb]{1,0,1}}%
      \expandafter\def\csname LT4\endcsname{\color[rgb]{0,1,1}}%
      \expandafter\def\csname LT5\endcsname{\color[rgb]{1,1,0}}%
      \expandafter\def\csname LT6\endcsname{\color[rgb]{0,0,0}}%
      \expandafter\def\csname LT7\endcsname{\color[rgb]{1,0.3,0}}%
      \expandafter\def\csname LT8\endcsname{\color[rgb]{0.5,0.5,0.5}}%
    \else
      \def\colorrgb#1{\color{black}}%
      \def\colorgray#1{\color[gray]{#1}}%
      \expandafter\def\csname LTw\endcsname{\color{white}}%
      \expandafter\def\csname LTb\endcsname{\color{black}}%
      \expandafter\def\csname LTa\endcsname{\color{black}}%
      \expandafter\def\csname LT0\endcsname{\color{black}}%
      \expandafter\def\csname LT1\endcsname{\color{black}}%
      \expandafter\def\csname LT2\endcsname{\color{black}}%
      \expandafter\def\csname LT3\endcsname{\color{black}}%
      \expandafter\def\csname LT4\endcsname{\color{black}}%
      \expandafter\def\csname LT5\endcsname{\color{black}}%
      \expandafter\def\csname LT6\endcsname{\color{black}}%
      \expandafter\def\csname LT7\endcsname{\color{black}}%
      \expandafter\def\csname LT8\endcsname{\color{black}}%
    \fi
  \fi
  \setlength{\unitlength}{0.0500bp}%
  \begin{picture}(11520.00,9072.00)%
    \gplgaddtomacro\gplbacktext{%
      \csname LTb\endcsname%
      \put(1078,5240){\makebox(0,0)[r]{\strut{}-0.01}}%
      \put(1078,6132){\makebox(0,0)[r]{\strut{} 0}}%
      \put(1078,7023){\makebox(0,0)[r]{\strut{} 0.01}}%
      \put(1078,7915){\makebox(0,0)[r]{\strut{} 0.02}}%
      \put(1078,8806){\makebox(0,0)[r]{\strut{} 0.03}}%
      \put(1210,5020){\makebox(0,0){\strut{}$10^{-4}$}}%
      \put(2594,5020){\makebox(0,0){\strut{}$10^{-2}$}}%
      \put(3979,5020){\makebox(0,0){\strut{}$10^{0}$}}%
      \put(5363,5020){\makebox(0,0){\strut{}$10^{2}$}}%
      \put(176,7023){\rotatebox{-270}{\makebox(0,0){\strut{}$\left. \Delta\mu_{12}^{tr} \middle/ \sqrt{\mu_{\perp}^t \mu_{\perp}^r} \right. = \left. \Delta\mu_{21}^{rt} \middle/ \sqrt{\mu_{\perp}^t \mu_{\perp}^r} \right.$}}}%
      \put(3286,4690){\makebox(0,0){\strut{}$\beta$}}%
      \put(156,8806){\makebox(0,0){\large $a)$}}%
    }%
    \gplgaddtomacro\gplfronttext{%
    }%
    \gplgaddtomacro\gplbacktext{%
      \csname LTb\endcsname%
      \put(6970,5240){\makebox(0,0)[r]{\strut{}-0.005}}%
      \put(6970,6132){\makebox(0,0)[r]{\strut{} 0}}%
      \put(6970,7023){\makebox(0,0)[r]{\strut{} 0.005}}%
      \put(6970,7915){\makebox(0,0)[r]{\strut{} 0.01}}%
      \put(6970,8806){\makebox(0,0)[r]{\strut{} 0.015}}%
      \put(7102,5020){\makebox(0,0){\strut{}$10^{-4}$}}%
      \put(8442,5020){\makebox(0,0){\strut{}$10^{-2}$}}%
      \put(9782,5020){\makebox(0,0){\strut{}$10^{0}$}}%
      \put(11122,5020){\makebox(0,0){\strut{}$10^{2}$}}%
      \put(5936,7023){\rotatebox{-270}{\makebox(0,0){\strut{}$\left. \Delta\mu_{23}^{tr} \middle/ \sqrt{\mu_{\perp}^t \mu_{\perp}^r} \right. = \left. \Delta\mu_{32}^{rt} \middle/ \sqrt{\mu_{\perp}^t \mu_{\perp}^r} \right.$}}}%
      \put(9112,4690){\makebox(0,0){\strut{}$\beta$}}%
      \put(5964,8806){\makebox(0,0){\large $b)$}}%
    }%
    \gplgaddtomacro\gplfronttext{%
    }%
    \gplgaddtomacro\gplbacktext{%
      \csname LTb\endcsname%
      \put(3958,704){\makebox(0,0)[r]{\strut{}-0.03}}%
      \put(3958,1596){\makebox(0,0)[r]{\strut{}-0.02}}%
      \put(3958,2488){\makebox(0,0)[r]{\strut{}-0.01}}%
      \put(3958,3379){\makebox(0,0)[r]{\strut{} 0}}%
      \put(3958,4271){\makebox(0,0)[r]{\strut{} 0.01}}%
      \put(4090,484){\makebox(0,0){\strut{}$10^{-4}$}}%
      \put(5474,484){\makebox(0,0){\strut{}$10^{-2}$}}%
      \put(6858,484){\makebox(0,0){\strut{}$10^{0}$}}%
      \put(8242,484){\makebox(0,0){\strut{}$10^{2}$}}%
      \put(3056,2487){\rotatebox{-270}{\makebox(0,0){\strut{}$\left. \Delta\mu_{32}^{tr} \middle/ \sqrt{\mu_{\perp}^t \mu_{\perp}^r} \right. = \left. \Delta\mu_{23}^{rt} \middle/ \sqrt{\mu_{\perp}^t \mu_{\perp}^r} \right.$}}}%
      \put(6166,154){\makebox(0,0){\strut{}$\beta$}}%
      \put(3036,4271){\makebox(0,0){\large $c)$}}%
    }%
    \gplgaddtomacro\gplfronttext{%
    }%
    \gplbacktext
    \put(0,0){\includegraphics{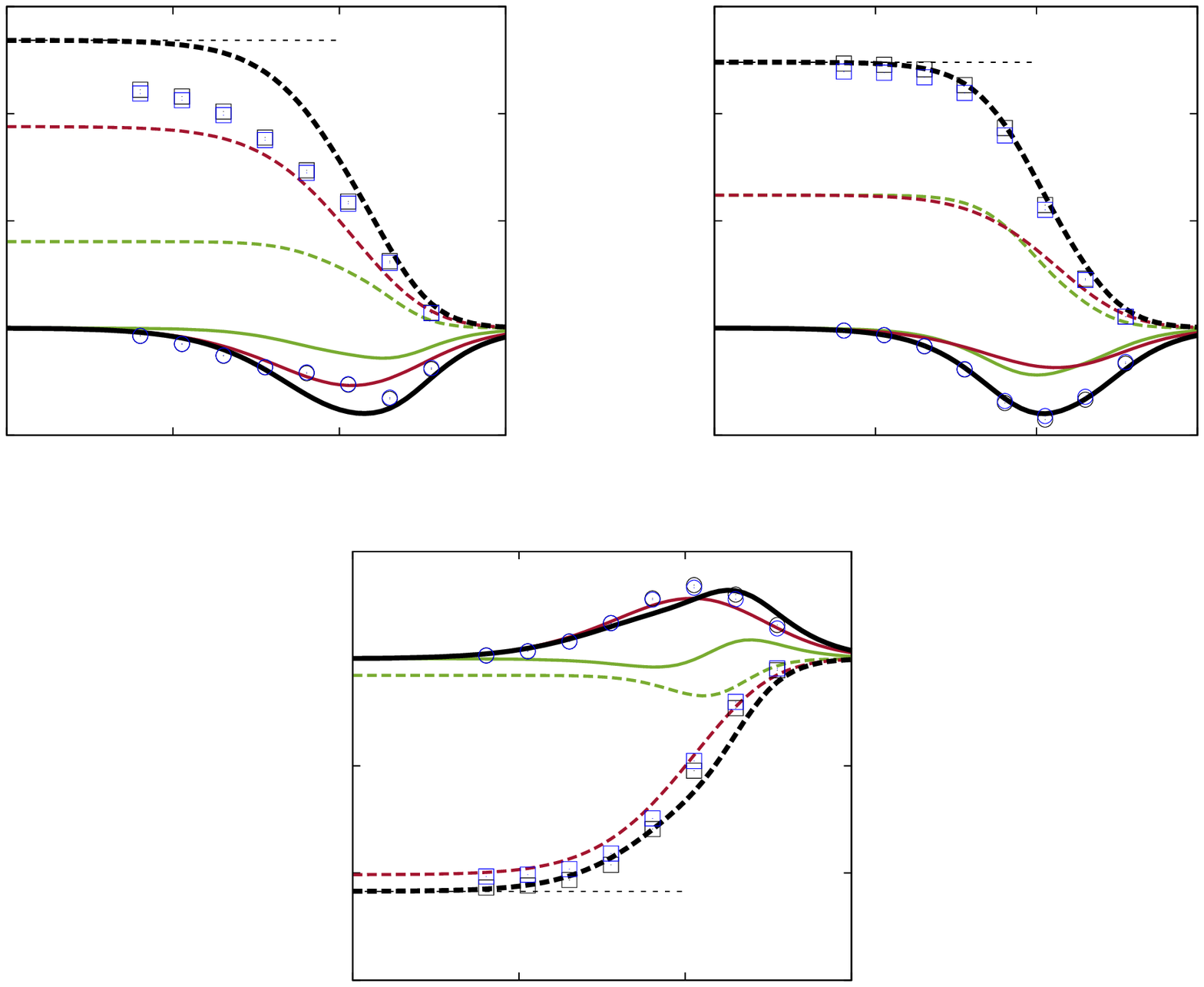}}%
    \gplfronttext
  \end{picture}%
\endgroup

%% file: Pics/DeltaMu_Rot.tex
\begingroup
  \makeatletter
  \providecommand\color[2][]{%
    \GenericError{(gnuplot) \space\space\space\@spaces}{%
      Package color not loaded in conjunction with
      terminal option `colourtext'%
    }{See the gnuplot documentation for explanation.%
    }{Either use 'blacktext' in gnuplot or load the package
      color.sty in LaTeX.}%
    \renewcommand\color[2][]{}%
  }%
  \providecommand\includegraphics[2][]{%
    \GenericError{(gnuplot) \space\space\space\@spaces}{%
      Package graphicx or graphics not loaded%
    }{See the gnuplot documentation for explanation.%
    }{The gnuplot epslatex terminal needs graphicx.sty or graphics.sty.}%
    \renewcommand\includegraphics[2][]{}%
  }%
  \providecommand\rotatebox[2]{#2}%
  \@ifundefined{ifGPcolor}{%
    \newif\ifGPcolor
    \GPcolorfalse
  }{}%
  \@ifundefined{ifGPblacktext}{%
    \newif\ifGPblacktext
    \GPblacktexttrue
  }{}%
  \let\gplgaddtomacro\g@addto@macro
  \gdef\gplbacktext{}%
  \gdef\gplfronttext{}%
  \makeatother
  \ifGPblacktext
    \def\colorrgb#1{}%
    \def\colorgray#1{}%
  \else
    \ifGPcolor
      \def\colorrgb#1{\color[rgb]{#1}}%
      \def\colorgray#1{\color[gray]{#1}}%
      \expandafter\def\csname LTw\endcsname{\color{white}}%
      \expandafter\def\csname LTb\endcsname{\color{black}}%
      \expandafter\def\csname LTa\endcsname{\color{black}}%
      \expandafter\def\csname LT0\endcsname{\color[rgb]{1,0,0}}%
      \expandafter\def\csname LT1\endcsname{\color[rgb]{0,1,0}}%
      \expandafter\def\csname LT2\endcsname{\color[rgb]{0,0,1}}%
      \expandafter\def\csname LT3\endcsname{\color[rgb]{1,0,1}}%
      \expandafter\def\csname LT4\endcsname{\color[rgb]{0,1,1}}%
      \expandafter\def\csname LT5\endcsname{\color[rgb]{1,1,0}}%
      \expandafter\def\csname LT6\endcsname{\color[rgb]{0,0,0}}%
      \expandafter\def\csname LT7\endcsname{\color[rgb]{1,0.3,0}}%
      \expandafter\def\csname LT8\endcsname{\color[rgb]{0.5,0.5,0.5}}%
    \else
      \def\colorrgb#1{\color{black}}%
      \def\colorgray#1{\color[gray]{#1}}%
      \expandafter\def\csname LTw\endcsname{\color{white}}%
      \expandafter\def\csname LTb\endcsname{\color{black}}%
      \expandafter\def\csname LTa\endcsname{\color{black}}%
      \expandafter\def\csname LT0\endcsname{\color{black}}%
      \expandafter\def\csname LT1\endcsname{\color{black}}%
      \expandafter\def\csname LT2\endcsname{\color{black}}%
      \expandafter\def\csname LT3\endcsname{\color{black}}%
      \expandafter\def\csname LT4\endcsname{\color{black}}%
      \expandafter\def\csname LT5\endcsname{\color{black}}%
      \expandafter\def\csname LT6\endcsname{\color{black}}%
      \expandafter\def\csname LT7\endcsname{\color{black}}%
      \expandafter\def\csname LT8\endcsname{\color{black}}%
    \fi
  \fi
  \setlength{\unitlength}{0.0500bp}%
  \begin{picture}(11520.00,9072.00)%
    \gplgaddtomacro\gplbacktext{%
      \csname LTb\endcsname%
      \put(1210,5240){\makebox(0,0)[r]{\strut{}-0.015}}%
      \put(1210,6132){\makebox(0,0)[r]{\strut{}-0.01}}%
      \put(1210,7023){\makebox(0,0)[r]{\strut{}-0.005}}%
      \put(1210,7915){\makebox(0,0)[r]{\strut{} 0}}%
      \put(1210,8806){\makebox(0,0)[r]{\strut{} 0.005}}%
      \put(1342,5020){\makebox(0,0){\strut{}$10^{-4}$}}%
      \put(2682,5020){\makebox(0,0){\strut{}$10^{-2}$}}%
      \put(4023,5020){\makebox(0,0){\strut{}$10^{0}$}}%
      \put(5363,5020){\makebox(0,0){\strut{}$10^{2}$}}%
      \put(176,7023){\rotatebox{-270}{\makebox(0,0){\strut{}$\left. \Delta\mu_{33}^{rr} \middle/ \mu_\perp^r \right.$}}}%
      \put(3352,4690){\makebox(0,0){\strut{}$\beta$}}%
      \put(203,8806){\makebox(0,0){\large $a)$}}%
    }%
    \gplgaddtomacro\gplfronttext{%
    }%
    \gplgaddtomacro\gplbacktext{%
      \csname LTb\endcsname%
      \put(7102,5240){\makebox(0,0)[r]{\strut{}-0.0015}}%
      \put(7102,6132){\makebox(0,0)[r]{\strut{}-0.001}}%
      \put(7102,7023){\makebox(0,0)[r]{\strut{}-0.0005}}%
      \put(7102,7915){\makebox(0,0)[r]{\strut{} 0}}%
      \put(7102,8806){\makebox(0,0)[r]{\strut{} 0.0005}}%
      \put(7234,5020){\makebox(0,0){\strut{}$10^{-4}$}}%
      \put(8530,5020){\makebox(0,0){\strut{}$10^{-2}$}}%
      \put(9826,5020){\makebox(0,0){\strut{}$10^{0}$}}%
      \put(11122,5020){\makebox(0,0){\strut{}$10^{2}$}}%
      \put(5936,7023){\rotatebox{-270}{\makebox(0,0){\strut{}$\left. \Delta\mu_{13}^{rr} \middle/ \sqrt{\mu_{\parallel}^r \mu_{\perp}^r} \right. = \left. \Delta\mu_{31}^{rr} \middle/ \sqrt{\mu_{\parallel}^r \mu_{\perp}^r} \right.$~~~~}}}%
      \put(9178,4690){\makebox(0,0){\strut{}$\beta$}}%
      \put(5938,8806){\makebox(0,0){\large $b)$}}%
    }%
    \gplgaddtomacro\gplfronttext{%
    }%
    \gplgaddtomacro\gplbacktext{%
      \csname LTb\endcsname%
      \put(1210,704){\makebox(0,0)[r]{\strut{}-0.008}}%
      \put(1210,1893){\makebox(0,0)[r]{\strut{}-0.004}}%
      \put(1210,3082){\makebox(0,0)[r]{\strut{} 0}}%
      \put(1210,4271){\makebox(0,0)[r]{\strut{} 0.004}}%
      \put(1342,484){\makebox(0,0){\strut{}$10^{-4}$}}%
      \put(2682,484){\makebox(0,0){\strut{}$10^{-2}$}}%
      \put(4023,484){\makebox(0,0){\strut{}$10^{0}$}}%
      \put(5363,484){\makebox(0,0){\strut{}$10^{2}$}}%
      \put(176,2487){\rotatebox{-270}{\makebox(0,0){\strut{}$\left. \Delta\mu_{11}^{rr} \middle/ \mu_\parallel^r \right.$}}}%
      \put(3352,154){\makebox(0,0){\strut{}$\beta$}}%
      \put(203,4271){\makebox(0,0){\large $c)$}}%
    }%
    \gplgaddtomacro\gplfronttext{%
    }%
    \gplgaddtomacro\gplbacktext{%
      \csname LTb\endcsname%
      \put(6838,704){\makebox(0,0)[r]{\strut{}-0.04}}%
      \put(6838,1417){\makebox(0,0)[r]{\strut{}-0.03}}%
      \put(6838,2131){\makebox(0,0)[r]{\strut{}-0.02}}%
      \put(6838,2844){\makebox(0,0)[r]{\strut{}-0.01}}%
      \put(6838,3558){\makebox(0,0)[r]{\strut{} 0}}%
      \put(6838,4271){\makebox(0,0)[r]{\strut{} 0.01}}%
      \put(6970,484){\makebox(0,0){\strut{}$10^{-4}$}}%
      \put(8354,484){\makebox(0,0){\strut{}$10^{-2}$}}%
      \put(9738,484){\makebox(0,0){\strut{}$10^{0}$}}%
      \put(11122,484){\makebox(0,0){\strut{}$10^{2}$}}%
      \put(5936,2487){\rotatebox{-270}{\makebox(0,0){\strut{}$\left. \Delta\mu_{22}^{rr} \middle/ \mu_\perp^r \right.$}}}%
      \put(9046,154){\makebox(0,0){\strut{}$\beta$}}%
      \put(5916,4271){\makebox(0,0){\large $d)$}}%
    }%
    \gplgaddtomacro\gplfronttext{%
    }%
    \gplbacktext
    \put(0,0){\includegraphics{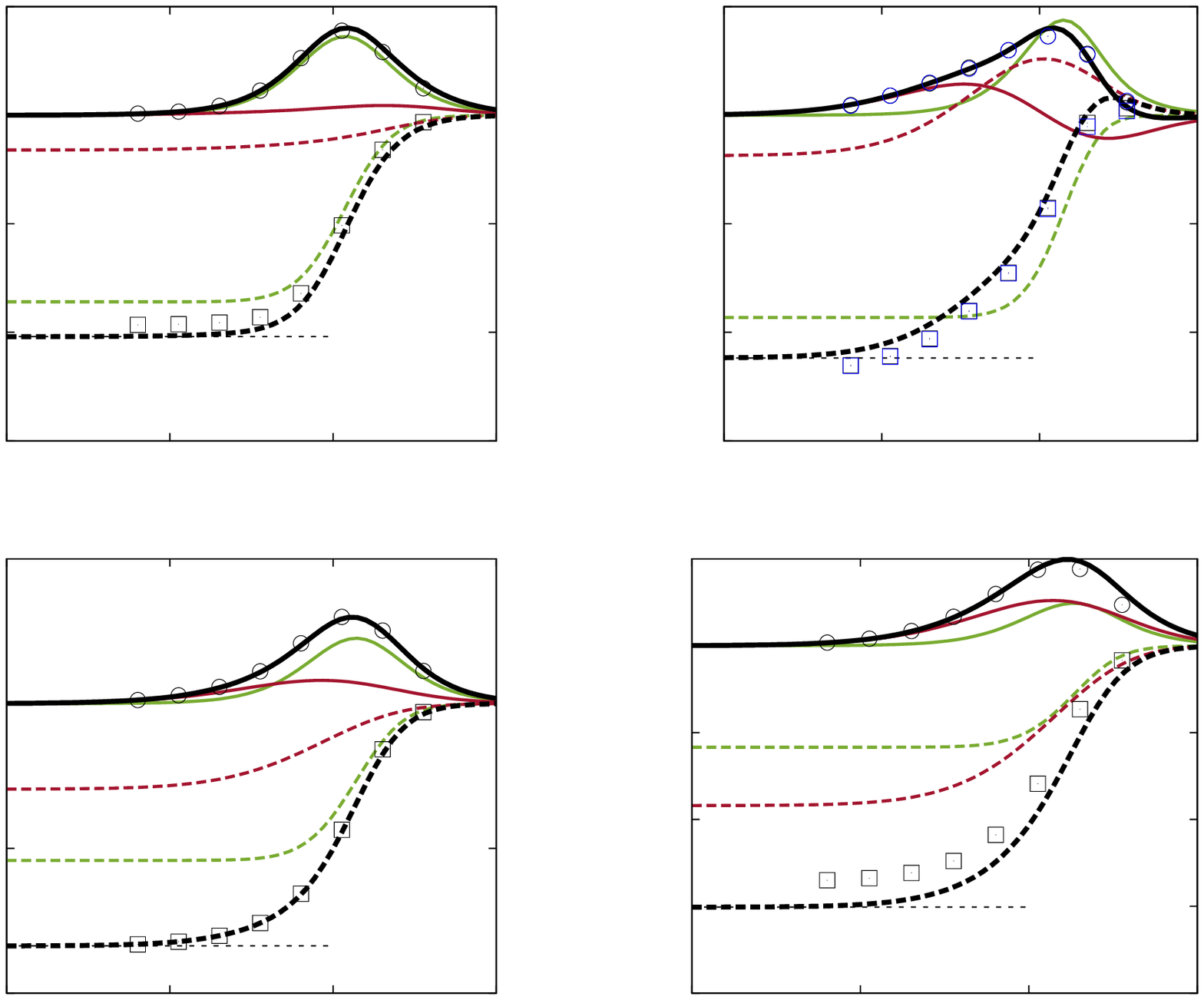}}%
    \gplfronttext
  \end{picture}%
\endgroup